\documentclass[twoside]{article}

\usepackage{PRIMEarxiv}

\usepackage[utf8]{inputenc} 
\usepackage[T1]{fontenc}    
\usepackage{hyperref}       
\usepackage{url}            
\usepackage{booktabs}       
\usepackage{amsfonts}       
\usepackage{nicefrac}       
\usepackage{microtype}      
\usepackage{lipsum}
\usepackage{fancyhdr}       
\usepackage{graphicx}       
\usepackage{tabularx} 
\usepackage{multirow}
\graphicspath{{images/}} 
\usepackage{subfigure}
\usepackage[colorinlistoftodos]{todonotes}
\usepackage{amsmath,amssymb}
 
\usepackage{listings}

\NewDocumentCommand{\halfcirc}{ O{} }{%
    \begin{tikzpicture}
        \fill[white] (0,0) circle (1.0ex); 
        \fill[black] (0,0) -- (270:1ex) arc (270:90:1ex) -- cycle; 
        \draw[black] (0,0) circle (1.0ex);
    \end{tikzpicture}
}
 
\NewDocumentCommand{\rot}{O{60} O{1em} m}{\makebox[#2][l]{\rotatebox{#1}{#3}}}%
\usepackage{pifont}

\newenvironment{example_continue}
{\addtocounter{example}{-1}\begin{example}{\textit{\textbf{{(continued)}}}}}
  {\end{example}}

\newtheorem{example}{Example}[section]  
\newtheorem{definition}{Definition}[section]

\usepackage{listings}
\definecolor{gray}{rgb}{0.4,0.4,0.4}
\definecolor{darkblue}{rgb}{0.0,0.0,0.6}
\definecolor{cyan}{rgb}{0.0,0.6,0.6}

\lstdefinelanguage{XML}
{
  morestring=[b]",
  morestring=[s]{>}{<},
  morecomment=[s]{<?}{?>},
  stringstyle=\color{black},
  identifierstyle=\color{darkblue},
  keywordstyle=\color{cyan},
  morekeywords={xmlns,version,type}
}

\title{A Deep Reinforcement Learning Approach for Security-Aware Service Acquisition in IoT}

\author{
  Marco Arazzi \\
  Department of Electrical, Computer and\\ Biomedical Engineering,\\ University of Pavia, Italy \\
  \texttt{marco.arazzi01@universitadipavia.it} \\
  \And
  Serena Nicolazzo\\
  Department of Computer Science,\\
  University of Milan, Italy \\
  \texttt{serena.nicolazzo@unimi.it} \\
  \And
  Antonino Nocera \\
  Department of Electrical, Computer and\\ Biomedical Engineering,\\ University of Pavia, Italy \\
  \texttt{antonino.nocera@unipv.it} \\
}

\begin{document}
\maketitle

\begin{abstract}
The novel Internet of Things (IoT) paradigm is composed of a growing number of heterogeneous smart objects and services that are transforming architectures and applications, increasing systems' complexity, and the need for reliability and autonomy. 
   In this context, both smart objects and services are often provided by third parties which do not give full transparency regarding the security and privacy of the features offered. Although machine-based Service Level Agreements (SLA) have been recently leveraged to establish and share policies in Cloud-based scenarios, and also in the IoT context, the issue of making end users aware of the overall system security levels and the fulfillment of their privacy requirements through the provision of the requested service remains a challenging task. 
   To tackle this problem, we propose a complete framework that defines suitable levels of privacy and security requirements in the acquisition of services in IoT, according to the user needs. Through the use of a Reinforcement Learning based solution, a user agent, inside the environment, is trained to choose the best smart objects granting access to the target services. Moreover, the solution is designed to guarantee deadline requirements and user security and privacy needs. Finally, to evaluate the correctness and the performance of the proposed approach we illustrate an extensive experimental analysis.
\end{abstract}


\section{Introduction}

The Internet of Things (IoT, hereafter) is a modern scenario characterized by interconnected smart objects that offer disparate services allowing autonomy and transparency to end-users. The intrinsic heterogeneity and multiplicity of devices composing the IoT and the connections they create, make it possible for an actor of the system (i.e., a human or an object on her/his behalf) to access services from different objects in a flexible way. The higher the number of objects and services provided, the greater the complexity of the system architecture design and objects' orchestration. 

Furthermore, in this setting, the attack surface is widening faster, and possible cybersecurity threats are evolving as well, targeting both data and object integrity \cite{ferretti2021h2o}. The existence of weak objects acting as a possible backdoor, the lack of device updates, the presence of inappropriate security practices, or the absence of them could act as a deterrent to the further evolution of IoT applications. This scenario is also exacerbated by the fact that, from a user side, a complete awareness of the security and privacy controls 
implemented by the different devices or brands is difficult to achieve \cite{tawalbeh2020iot,arazzi2024novel,arazzi2023fully}.
The lack of transparency can produce unclear situations fostering the so-called ``privacy paradox", a recent phenomenon referring to a discrepancy between a person's attitude towards privacy and her/his actual behavior. Indeed, when interacting with IoT or Internet-based services, users claim to be seriously concerned about their privacy and dismayed at the possibility that their data may be used for a purpose other than that for which it was collected, but, in fact, not performing any conscious choices, they disclose their sensitive information anyway to benefit from a given service \cite{barth2019putting}. Interestingly this 
behavior is based on a general psychological process that takes place during decision-making: i.e., during the user risk-benefit calculation phase before the action of benefiting a service, concerns are overridden by factors such as desirability of the application, time constraints, or promised gratifications. Despite this, a foreboding feeling persists in the user leading to a generally wary attitude towards the credibility of privacy protection mechanisms \cite{barth2017privacy}.

To solve this issue, Security Service Level Agreements (SecSLAs) and Privacy Level Agreements (PLAs) have been recently developed to formalize and assess the security and privacy levels for different smart devices interacting with a Cloud application \cite{rios2022security,casola2018security}. However, one of the main drawbacks of this practice is the lack of a holistic approach providing standard security protocols and policies for all the heterogeneous objects forming an IoT network. Moreover, the use of Service Level Agreements (SLAs, hereafter) can lead to requesting irrelevant permissions from users, creating loosely defining permissions, or misusing them.

A possible solution to the above-mentioned drawbacks of using SLAs can rely on user empowerment, to make users conscious and even protagonists of the distribution and sharing of their data.
Actually, in the current scenario, users (or objects on their behalf) interacting with different devices providing services have no tools to make a suitable choice according to the offered system security or privacy levels.

To tackle the challenge of empowering users with the possibility of acquiring services according to their requirements and overcoming the limitations of the classical use of SLA, in this paper, we design a system that allows users to specify their security and privacy needs before interacting with the IoT environment. Moreover, due to the complexity and high dimensionality of the problem, our system is based on a heuristics strategy leveraging a Deep Reinforcement Learning (DRL, hereafter) approach to train an agent to take actions on behalf of its user. The agent's goal is to select, for every user request, the best service providers based on the expressed requirements.

In our IoT scenario, at first, a user has to express her/his needs in terms of privacy and security requirements and after that, a smart object playing the role of user agent will interact and sense the environment to collect relevant information and take the sub-optimal actions to satisfy her/his owner needs. In particular, while the agent is moving inside the environment and is interacting with other peers, it is called to provide the needed actions, through service provisioning available in the environment, and also accomplish the user security and privacy constraints. For this purpose, a tailored reward function is created which, depending on the model parameters, rewards an agent if it performs the required actions by accepting services offered by peers it just met, or by peers met before and saved in its list of contacts. In some cases, also performing no interaction at all can be rewarded if, in doing so, the advantage in terms of security and privacy protection is more valuable than obtaining the desired service.
Interestingly, during its movements, the agent builds a list of contacts with nodes met in the environment in a given time window. Through our experiments, we show that this mechanism allows us to obtain a higher percentage of acquired services as well as the best fitting of user privacy requirements. 

To the best of our knowledge, our approach 
is the first of this kind and it presents several novelties and scientific contributions, such as:

\begin{itemize}
    \item It empowers users with the possibility to decide their level of security and privacy needs, making them more proficient and conscious of what they disclose and increasing their feeling of safety when it comes to smart objects and IoT technologies. 
    \item It rigorously formalizes the problem of choosing the right service provider based on privacy requirements in the environment through a decision-making problem framed into a Deep Reinforcement Learning (DRL) solution. 
    \item Finally, it takes advantage of the peculiarities of this DRL-based strategy allowing an agent to learn from experience, by moving in a multi-dimensional and complex environment and reacting to possible changes. This makes the approach more efficient and suitable for an IoT scenario \cite{chen2021deep}.
\end{itemize}

The outline of this paper is as follows. Section \ref{sec:UseCase} is devoted to the description of a reference scenario that could benefit from our proposed solution. In Section \ref{sec:RelatedWork}, we examine papers of the literature related to our approach. In Section \ref{sec:Description}, we give a general overview of our reference IoT model and describe the proposed framework in detail. Section \ref{sec:drl} focuses on the DRL solution of our approach. In Section \ref{sec:Experiments}, we present the set of experiments carried out to test our approach and show its performance. Finally, Section \ref{sec:Conclusion} discusses interesting leads as future work and draws our conclusions.

\section{Use case scenario}
\label{sec:UseCase}

As discussed above, IoT is becoming a key element of disparate and variegated application domains, mainly due to the pervasiveness and heterogeneity of the involved ``smart'' devices.
This phenomenon is so important that novel paradigms, such as Society 5.0 and Industry 5.0 \cite{fujii2018consideration,maddikunta2022industry}, assume the integration of people's daily lives with smart technology and IoT in their manifesto, directly. 
Although such paradigms are somehow at an initial stage, real-life examples of such an integration are already part of people's lives.
In this paper, we focus on Smart Cities \cite{theodoridis2013developing} as a reference scenario for our solution. In such ecosystems, IoT plays a crucial role in the enhancement of efficiency, sustainability, and overall quality of resident's and tourist's lives. An ever-growing number of cities are investing in the development and improvement of services driven by IoT technologies. Meaningful examples are Santander\footnote{https://smartsantander.eu/}, Oulu\footnote{https://nscn.eu/OuluSmartCity} and Dubai\footnote{https://www.digitaldubai.ae/initiatives}.In these cities, several solutions have been developed leveraging lower-level services provided by IoT devices; for instance, to name a few:

\begin{itemize}
    \item Traffic and parking management. Leveraging real-time data from sensors in traffic lights, roads, and vehicles, the underlying IoT can help optimize traffic flow and parking.
    \item Environment monitoring. Hundreds of air pollution sensors, noise sensors, and objects measuring the water quality or the temperature have been installed on lamp posts, public transport buses, and police cars.
    \item Augmented reality. This facility enables the tagging of points of interest (POIs) in the city, such as historical sites, parks, public services, and shops. When inside these areas of interest, people can receive notification of events, exhibitions, or offers.
    \item Healthcare. IoT can support public health efforts by monitoring air quality, tracking disease outbreaks, and even enabling remote patient monitoring.
    \item Public Safety. IoT can enhance public safety through video surveillance, gunshot detection, and other security measures that can provide real-time alerts and data to law enforcement, also in case of emergency. In particular, thanks to the possibility for the users to utilize their own wearable devices and send physical sensing information to other smart objects (e.g. GPS coordinates, compass, environmental data such as noise, temperature, etc.), it is possible to fine-tune security notifications for specific types of events (traffic jams, car
    accidents, concerts, protests, etc.).
\end{itemize}

\begin{figure*}[ht]
    \centering
    \includegraphics[width=0.7\textwidth]{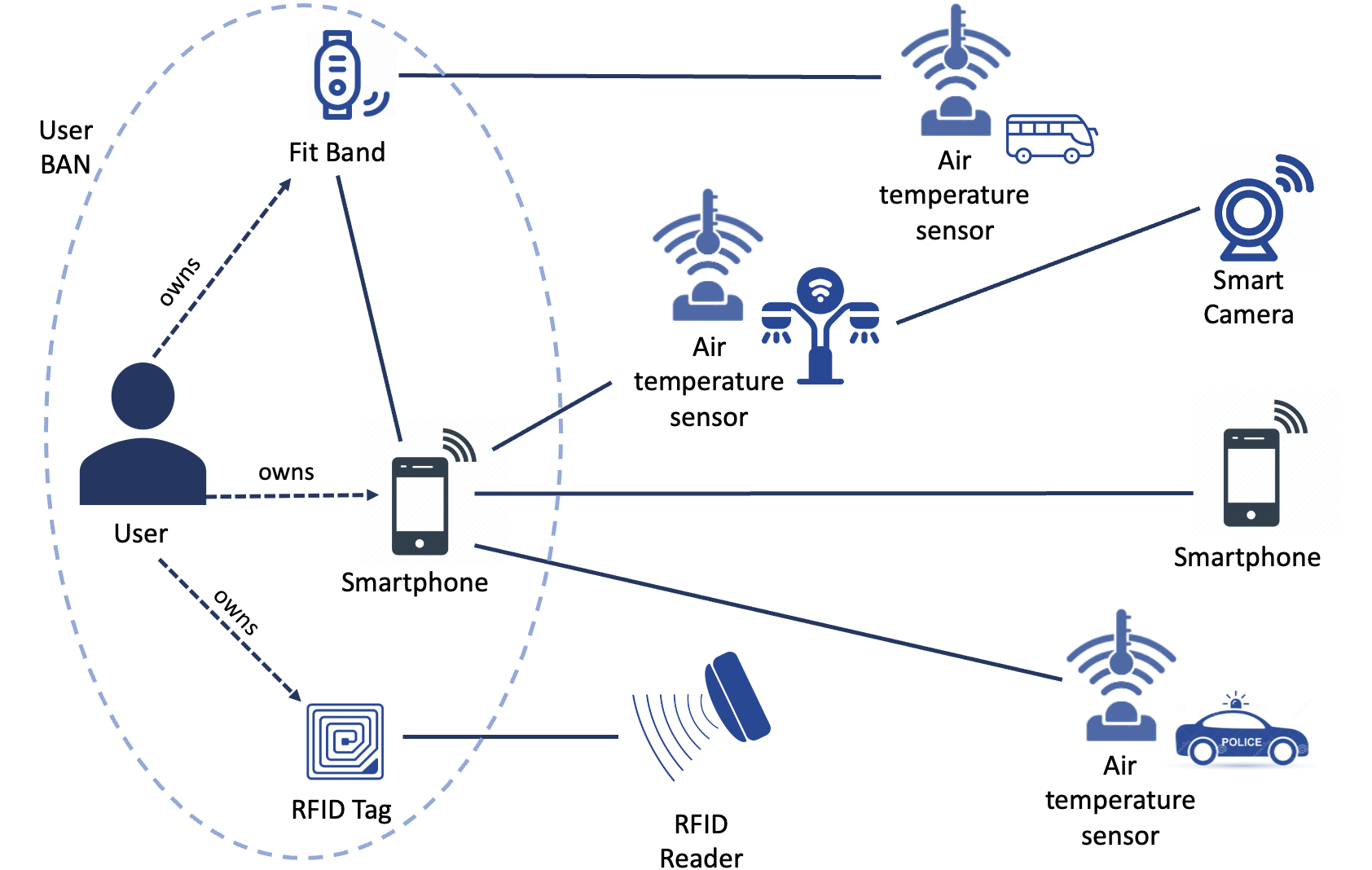}
    \caption{IoT reference scenario  \label{fig:scenarioIoT}}
\end{figure*}

In general, users equipped with personal devices and moving into a smart city aim at accessing different services provided by surrounding smart things. 
As shown in the example reported in Figure \ref{fig:scenarioIoT}, while walking or jogging in the city, a user can wear a Fit Band, she/he can have a smartphone with her/him, an RFID Tag, and so forth, all linked together in the user's Body Area Network (BAN). These smart devices are supposed to perform actions on behalf of their owners, through the services available in the surrounding environment. For instance, the Fit Band should show the updated local environment temperature and/or the instantaneous pollution level, throughout the whole user travel route. Multiples providers offering this information are available in a Smart City network, e.g., sensors installed on lamp posts, public transport buses, police cars, Points of Interest (e.g., shops or drugstores), or services offered even by other users' devices (e.g., via community-oriented applications for fitness).
In this complex scenario, identifying the best provider for a target service in a set of intrigued and variegated potential candidates appears an extremely challenging task.
This holds even more if we consider that such a choice should be strongly bounded by user-expressed requirements and time constraints.

\section{Related Work}
\label{sec:RelatedWork}

For the past few years, the interest in Artificial Intelligence (AI, hereafter) has grown at a very fast pace, with numerous algorithms rapidly developed and exploited in a wide range of application scenarios.

A field of research that has recently gained much more attention from both academic and industrial communities is Deep Reinforcement Learning (DRL)\cite{mnih2013playing}. This family of techniques, which combines convolutional neural networks and Q-learning, has been used to solve complex sequential decision-making problems in an effective way\cite{theate2021application}.
IoT is a new scenario that most benefits the peculiarities of DRL, such as its dynamic decision-making mechanisms and the possibility to deal with incomplete information related to their environments. Indeed, numerous
research proposals have applied Reinforcement Learning and DRL techniques to overcome the
difficulties in the monitoring and management of IoT networks due to the growing number of their devices, the increasing complexity of systems, and
the large volume of generated data \cite{frikha2021reinforcement}. In particular, recent DRL techniques have replaced traditional RL-based approaches to make the learning and the decision operations more efficient, the computation less complex and to reduce the storage space, thanks to the combination of RL and Deep Learning \cite{lei2020deep,nguyen2020deep}. For instance, the work presented in \cite{liu2019deep} deals with an IoT network dynamic clustering solution in edge computing using DRL. The authors aim to optimize the data aggregation at the edge server located in the cluster. Another IoT application of DRL is illustrated in \cite{liang2020deep} where a DRL-based detection
algorithm for virtual IP watermarks is proposed to process
the ownership information of the IP circuit resource.

Few works adopt DRL and RL for security in IoT. For instance, \cite{mohammed2020ubipriseq} illustrates the functioning of the UbiPriSEQ framework leveraging the DRL technique for optimizing QoS, energy efficiency preserving security, and privacy in 5G networks.
Instead, the work presented in \cite{al2020blockchain} introduced a blockchain-based solution in the context of Industrial IoT (IIoT) service delivery to cloud users. It aims to autonomously create user-defined applications to authenticate and deliver composite IoT services. The composition process is based on an RL technique to construct secure and reliable composition paths. Always in the same IIoT context, the authors of \cite{xu2022c} propose a framework called C-fDRL to
provide federated DRL to maintain the context-aware privacy of the task offloading in the Cloud environment.

The above contexts (i.e., 5G and Industrial IoT) are very far from the ones we treat. The algorithmic solution presented in our proposal leverages the Deep Q-Network (DQN) algorithm, which has been adapted to the particular sequential decision-making problem of finding the right thread-off between user privacy requirements and the time of fulfillment of a service needed by the user. To do so, in our framework, we exploit the potentialities of Service Level Agreements (SLA, hereafter).

In recent years, research on SLA composition has been a growing topic, especially in the context of the interaction with Cloud providers and IoT \cite{di2016supporting,rios2022security}.
Notable proposals include ontology-based techniques \cite{liu2012sla} and functional service composition methods \cite{zappatore2015sla}. Nevertheless, these and similar previous works focus on SLAs for functional and performance requirements (e.g. response time, MTTR, etc.) which cannot be easily remapped to security and privacy requirements.

Whereas, the approach of \cite{di2016supporting} designs a CSP-based automated framework for composing SLAs and it exploits both application
requirements and dependencies among the characteristics of
a provider. Hence, the SLA between the subject
and a Cloud Provider (CP), rather than being based on a pre-defined
model produced by the CP is created by considering all specific requirements that characterize the application. Moreover, the approach of \cite{rios2022security} specifically focuses on SecSLAs and PLAs of Cloud-based IoT applications and presents a methodology to compose them on top of standard controls. Also, the work presented in \cite{rios2019service} is related to this context and describes a framework to support Cloud consumers in designing, deploying, and operating multi-cloud systems that include the
necessary privacy and security controls. In particular, it designs a solution to SLA-based privacy and
security assurance for multi-cloud applications. 
Finally, the proposals in \cite{di2017supporting} deal with a consensus-based approach for accommodating contrasting requirements from multiple applications.
All the approaches presented above focus on SLA composition, our proposal, instead adopts a different perspective, indeed, it leverages SecSLAs and PLAs to derive some security level of the required service and use this to choose the most suitable service according to the implemented DRL algorithm.

\section{Description of Our Approach}
\label{sec:Description}

In the next subsections, we provide a general overview of our solution along with its underlying model.

\subsection{The underlying IoT Model}
\label{sub:modelIoT}

This section details the architectural design of our DRL-based approach. In particular, we will describe the system actors and how they interact with each other during the model learning and evaluation processes.  

We consider a reference scenario characterized by IoT nodes interacting with each other in a complex environment in which some of them have the role of service providers and others are service users.
The latter nodes act on behalf of their owners, they move in the environment and they get in touch with service providers. For this reason, they are equipped with a software agent specifically designed to optimize their service acquisitions.
Our goal is to suitably design and train such an agent so that it is capable of choosing the right nodes providing services that better match the owner's requirements.
Such requirements include {\em operations} that the owners wish to obtain, along with their priority, security and privacy requirements, and time constraints. 

Table \ref{tab:SystemSymbols} reports all the abbreviations and symbols used throughout this paper.

\begin{table}[ht]
\scriptsize
\centering
\begin{tabular}{||l|l||}
\hline
\bf{Symbol} & \bf{Description}\\
\hline \hline
 $CP$ & Cloud Provider \\
 $DL$ & Deep Learning \\
 $DRL$ & Deep Reinforcement Learning \\
 $IoT$ & Internet of Things\\
 $RL$ & Reinforcement Learning \\
 $SLA$ & Service Level Agreement\\
 $SecSLA$ & Security Service Level Agreement\\
 $PLA$ & Privacy Level Agreement\\
 $at$ & Agent \\
 $\sigma$ & Service provided by a node\\
 $a$ & Attribute of a SLA\\
 $co$ & Constraint over an attribute $a$\\
 $l$ & Security label\\
 $p$ & Security property\\
 $c$ & Security class\\
 $\Phi$ & Service type \\
 $u$ & User survey \\
 $\mathcal{K}$ & Weighted security lattice\\
 $\rho$ & Path in a security lattice $\mathcal{K}$\\
\hline
\end{tabular}
\caption{Summary of the main symbols and acronyms used in the description of our approach section.\label{tab:SystemSymbols}}
\end{table}

As classically done in the literature, the considered IoT scenario can be modeled through an undirected graph $G= \langle V, E \rangle $, where $V= N \cup U$ is the set of vertices composed of nodes (referring to IoT objects) moving in the environment and the set of users $U$ owning (or administrating) them. As for the set of arcs $E= R \cup O$, it includes the set of edges representing relationships (encoding interactions) between pairs of nodes and also the set of ownership relationships between users and their nodes. 

In our scenario, a graph $G$ contains a set of $m$ agents $at=\{at_1,at_2,\dots,at_m\} \in A \subseteq N$. Note that, an agent is a device of $N$ with sufficient capabilities in terms of memory and computational strength to perform the more demanding tasks of our approach (e.g. the execution of DRL solutions); moreover, it is entrusted with a set of operations that it must carry out autonomously on behalf of the corresponding owner, during a single {\em life cycle}.
For a particular agent $at_i$, we can build an ego-network $\hat{G}_{at_i}=\langle \hat{N}_{at_i}, R \rangle$ as sub-graph of $G$, where $\hat{N}_{at_i}$ represents the group of nodes that an agent $at_i$ has been interacting with, and it is defined as $\hat{N}_{at_i}=\{n_j \in N : (at_i,n_j) \in E\}$. 
As typically happens for IoT devices, for computational capability reasons, the cardinality of the set $\hat{N}_{at_i}$, for each agent, is bounded to a maximum value. Indeed, the storing capabilities allow an agent to save only a finite number of past interactions.

In our referring scenario, each node of $N$ can act also as a {\em Service Provider}, and, in this case, it is in charge of delivering services or executing applications if requested. 
Generally speaking, the interaction between a service provider and an agent is regulated by Service Level Agreements.
Such agreements contain also a descriptor of the referring services, which may consist of multiple operations.
As for this last aspect, in our approach, we explicitly refer to the WSLA framework proposed by IBM \cite{ludwig2003web,bianco2008service}. Its language is based on XML
and defined as an XML schema. Primarily, the WSLA allows the creation of machine-readable SLAs for services implemented using Web services technology. 
An example of a WSLA specifying a ``GetTemperature'' operation inside a ``WeatherService'' service for an IoT provider is shown in Listing \ref{list:wsla}.

\begin{figure*}
\begin{lstlisting}[language=XML, caption=A portion of a SLA showing the mapping between services and operations, label=list:wsla]
        <ServiceDefinition name="WeatherService">
         . . .
         <Operation name="GetTemperature" xsi:type="WSDLSOAPOperationDescriptionType">
           <Metric name="Temperature" type="float" unit="Celsius">
             <Source>YMeasurement</Source>
             <Function xsi:type="Minus" resultType="float">
               <Operand> . . . </Operand>
             <Function xsi:type="Fahrenheit" resultType="float">
               <Operand> . . . </Operand>
             </Function>
           </Metric>
           . . .
         </Operation> 
         . . .
\end{lstlisting}
\end{figure*}

During the acquisition of a service (and the included operations), an agent can exchange sensitive information with the service provider, and several security features can be adopted to protect such interaction. For this reason, such specifications may also include privacy-related information, in the form of PLA, and security-related constraints, through SecSLA.
To provide instruction to the agent, in our scenario, its owner (the user of our solution) fills in a specific survey aiming at assessing her/his security and privacy protection requirements. 
Answers provided in the survey are, then, used to estimate configuration parameters for the agent, including a maximum {\em security loss} that the agent can cause during its interaction with the environment. 
As will be clearer in the following, our solution allows users to express their requirements simply by referring to high-level properties, in contrast to low-level configuration parameters usually SecSLAs and PLAs are composed of.

\subsection{Security Level Agreement mapping} 
\label{sub:pla}

As said above, the interaction between a service provider and an agent is usually regulated by SLA representing a contract between the two parties defined by standard ISO/IEC 20000-1 \cite{rovers1970iso}. Classically, an SLA is a written verbose document stating different functional/non-functional properties of a target service, which can be expressed through logical formulas involving the concepts from a common/shared ontology.  
However, several approaches promote the use of machine-readable SLAs to ease the specification, comparison, and monitoring of the controls, and above all improve the automation of SLA management \cite{paschke2005rbsla,kearney2010sla}.

Recently, two categories of SLA have been developed, known as Security Service Level Agreements (SecSLAs) and Privacy Level Agreements (PLAs), which focus on the formalization of the security and privacy specifications for different services \cite{rios2022security,casola2018security}.
Examples of SecSLA and PLA attributes can be the service response time, the adopted encryption function for the communications, the service location, etc. Clearly, to produce a service, a device can, on its side, interact with other entities or sub-systems (e.g., other peers, Cloud facilities, and so on), as such the resulting SecSLA/PLA can have different complexity levels.
Indeed, in a system where multiple distributed components are orchestrated to provide common services, the overall security depends on the protections offered by each component along with the underlying infrastructure.

In our scenario, we acknowledge that each service provisioning, as described by its SLA, is composed of a set of operations.
SecSLA/PLA attributes are, hence, referring to each operation included.

With that said, we can now provide a formal definition of a SecSLA/PLA describing its main components.
In the following, we will focus mainly on SecSLA, although the same reasoning can be applied to PLA as well.

Let $\mathcal{A}$ be the set of attributes. Each attribute $a \in \mathcal{A}$ takes values from its domain $\mathcal{D}(a)$. For instance, $\mathcal{D}(encr)$=\{SHA256, AES128, 3DES\} means that the possible possible encryption algorithms can be SHA256, AES128 or 3DES.
A constraint defined over an attribute restricts the values that the property represented by the attribute can assume. A constraint is formally defined as follows.

\begin{definition}{\bf{(Constraint)}}
Given an attribute $a \in \mathcal{A}$ in the set of attributes $\mathcal{A}$ and domain $\mathcal{D}(a)$ and a value $v \in \mathcal{D}(a)$, a constraint $co$ on $a$ can be expressed as $co : \langle a$ op $v \rangle$, with op $\in \{ =, \neq,<,\le,>,\ge\}$ a comparison operator.
\end{definition}

\noindent
For instance, $co_1:\langle resp\_time<5ms \rangle$ and $co_2:\langle encr=3DES \rangle$ model two conditions demanding that the service exhibits a response time lower than $5$ milliseconds ($co_1$) and uses 3DES as encryption method ($co_2$).

Starting from the definition of constraint, we can represent also the concepts of SecSLA, which can be seen as a set of constraints $\bar{\mathcal{CO}} = \{co_1,...,co_n\}$, whose enforcement must be guaranteed in the service provision.
Our framework is characterized by multiple devices each exposing a customized SecSLA composed of several constraints.

To detail our model we got inspired by the notion of {\em abstract security property} introduced in \cite{di2019security} and we apply it to our IoT scenario. In particular, we use this concept to {\em(i)} classify provided services, {\em(ii)} formulate user
protection requirements, and {\em(iii)} map user requirements in a proper security level or {\em security budget} available to the agent and that can be spent to complete user-required operations, through available service provisioning. This model is also used to quantify the security loss in accepting services provided by devices.
Abstract security properties model high-level concepts (such as confidentiality, integrity, and so forth) and associate them with a domain of labels, each corresponding to
the satisfaction level of the specific property.
The advantage of adopting such a model is that, in this way, the designed framework can handle the security requirements without exposing
specific attributes that contribute to
obtain the desired protection \cite{di2019security}.
Classically, abstract security properties are directly derived from the CIA triad (Confidentiality, Integrity, and Availability) but others can also be
added to this base depending on the desired granularity level \cite{nist2013}. 
Our model considers a generic set $\mathcal{P}$ of properties and, for each property $p_i \in \mathcal{P}$, the compliance level for a service can be represented through a set of high-level security labels $\mathcal{L}^{p_i}$. For instance, in case only the properties of the CIA triad are considered $\mathcal{P}$(C, I, A), each property can be associated with one of the following three security labels corresponding to the High (HC, HI, and HA), Medium (MC, MI, and MA), and Low (LC, LI, and LA) values. Again, such labels encode the extent to which the provided service is compliant with the reference property. The symbol $-$ is used to represent the case in which a property is not ensured at all.

As visible in Table \ref{tab:exampleSLA}, there exists a correspondence between a property $p_i \in \mathcal{P}$, a security label $l_j \in \mathcal{L}^{p_i}$ and the constraints of $\mathcal{\bar{CO}}$ defined in an SecSLA. 
Labels are defined in such a way that there exists a partial relationship among them, namely $\succ^p$. Moreover, we define a distance function $d(\cdot,\cdot):\mathcal{L}^{p_i}\times \mathcal{L}^{p_i} \rightarrow \mathbb{R}$ associating a real number to each pair of labels.

Table \ref{tab:exampleSLA} shows an example of the properties $\mathcal{P}=\{C, I, A\}$ using the labels $\mathcal{L}^C \cup \mathcal{L}^I \cup \mathcal{L}^A$. Here, $\mathcal{L}^C=\{HC, MC, LC, -\}$, with $HC \succ^C MC\succ^C LC\succ^C -$, is the set of labels associated with the property $C$;
$\mathcal{L}^I=\{HI, MI, LI, -\}$, with $HC \succ^I MC\succ^I LC\succ^I -$, is the set of labels associated with the property $I$; finally,
$\mathcal{L}^A=\{HA, MA, LA, -\}$, with $HC \succ^A MC\succ^A LC\succ^A -$, is the set of labels associated with the property $A $. 
This table also shows a possible mapping between security properties, labels, and constraints for a sample SecSLA. For instance, for the property $C$, the label $HC$ requires a strong authentication method for the object that (i.e., continuous authentication) and an encryption method used for data exchange equal to {\em AES-256}; whereas, the label $LC$ requires only a simple username and password scheme for authentication or the {\em AES-128} encryption method.

\begin{table*}[ht]
\centering
\caption{Security properties of the running example}
\label{tab:exampleSLA}
\begin{tabular}{ |c|c|c|c| } 
\hline
\textbf{Property $p$} & \textbf{Label $l$}& \textbf{Constraint $co$} \\
\hline
\multirow{3}{*}{C} & HC & $auth=continous \wedge enc=AES-256$ \\ 
& MC & $auth=double factor \vee enc=AES-256$ \\ 
& LC & $auth=simple \vee enc=AES-128$ \\ 
& $-$ & $true$ \\ 
\hline
\multirow{3}{*}{I} & HI & $Merkle Hash Tree$ \\ 
& MI & $Hash Chain$ \\ 
& LI & $Verification Object Insertion$ \\ 
& $-$ & $true$ \\
\hline
\multirow{3}{*}{A} & HA & $uptime>99.99\%$ \\ 
& MA & $uptime>99\%$ \\ 
& LA & $uptime>95\%$ \\ 
& $-$ & $true$ \\ 
\hline
\end{tabular}
\end{table*}

\begin{figure}[ht]
    \centering
    \includegraphics[width=0.45\textwidth]{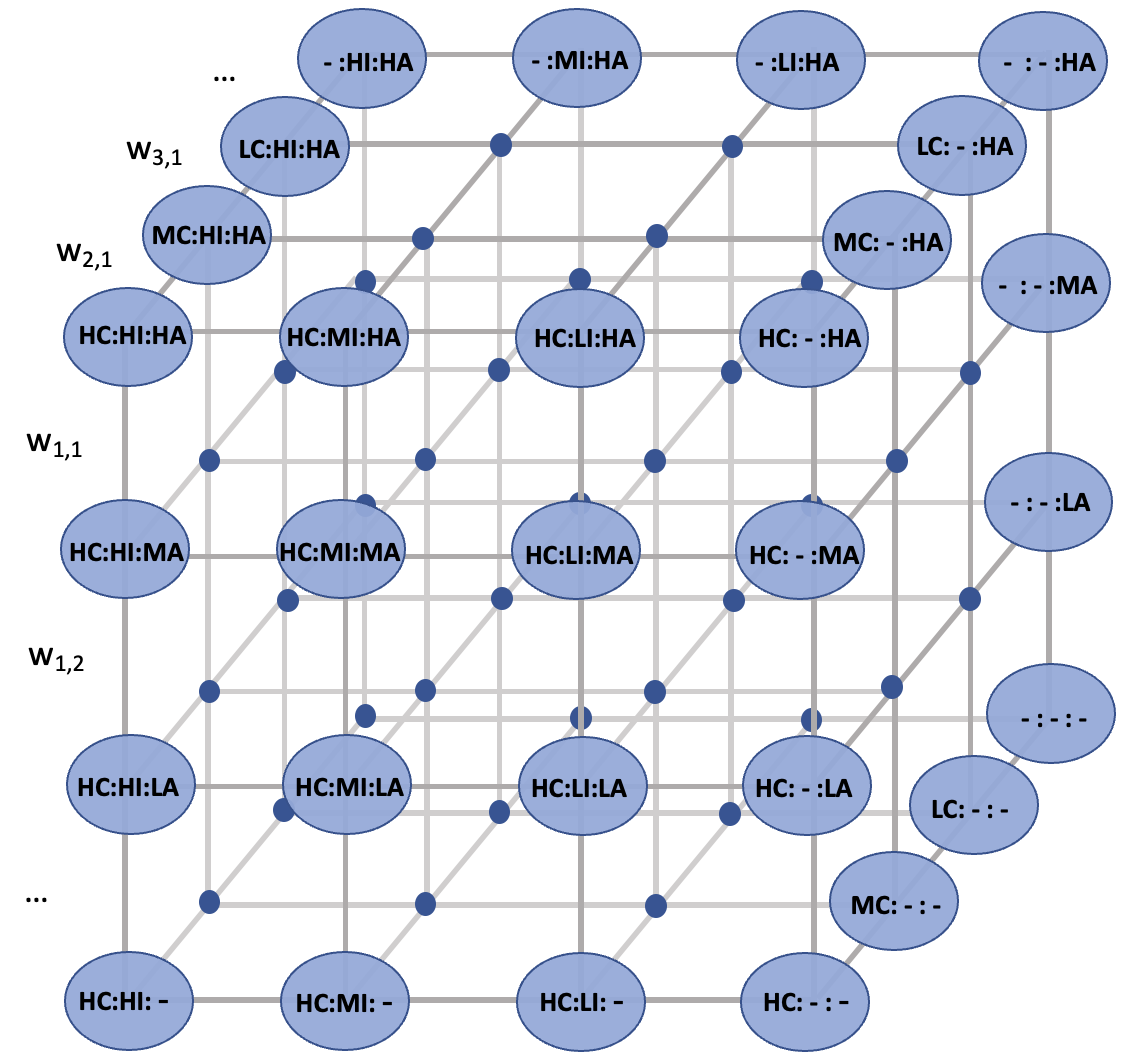}
    \caption{Weighted security lattice over the security properties. \label{fig:lattice}}
\end{figure}

Now given a mapping between properties, labels, and constraints, a SecSLA can be re-defined as a set of security properties along with the corresponding security labels. We can refer to this set as a {\em security class} $c_i$.  For instance, an example of a {\em security class} built for a SecSLA focusing on the properties of the CIA triad can be $c_i=[HC, HI, HA]$. 
The order relationship holding
within the labels of the different properties induces
a partial order relationship, $\succeq$, on the set of security classes $\mathcal{C}=\{c_1,\dots,c_n\}$. 
Moreover, the distance associated with each pair of labels in each ordered space can be used to estimate distances between the security classes of $\mathcal{C}$, thus obtaining a weighted context $(\mathcal{C},\mathcal{W})$, with $\mathcal{W}=(w_{i,j})_{1\leq i,j \leq |\mathcal{C}|}$ being a distance matrix.
This concept can be represented through a weighted
{\em security lattice} $\mathcal{K}=\langle (\mathcal{C}, \mathcal{W}) , \succeq \rangle$ over $\mathcal{P}$ as illustrated in Figure \ref{fig:lattice}.

A security class dominates another one if and only if the dominance relationship holds for each of its components. For instance, $[HC, HI, HA]\succeq[HC, MI, HA]$.
Moreover, the distance between two security classes is computed as the sum of the distances between the corresponding components.

We are now ready to define two important concepts used throughout this paper, namely {\em(i)} the security class of service and {\em(ii)} the security class of user requirements.

We say that a service $\sigma$ provided by a device satisfies a security label $l_j \in \mathcal{L}^{p_i}$ of a property $p_i$, denoted by $\sigma \models l_j$, if some of the constraints inside the SecSLA can be mapped to this security label.
Similarly, a service satisfies a security class $c_i$, denoted $\sigma \models c_i$, if all the constraints listed inside the SecSLA can be combined to map all the security labels defining $c_i$. 
Formally, we extend the definition of a security class introduced in \cite{di2019security}, as follows.

\begin{definition}{\bf{(Security class of a service)}}
    Given a service $\sigma$ and a weighted security lattice $\mathcal{K}$ over $\mathcal{P}$, the security class of $\sigma$, denoted $\lambda(\sigma)$, is the security class $c_i \in \mathcal{C}$ such that $\sigma \models c_i$, and $\nexists \ c_j \in \mathcal{C}$ with $c_j \succeq c_i$ and $\sigma \models c_j$.
\end{definition}

\noindent

We now define the specification of user protection
requirements established for a user through a survey filled in when she/he joins our system. 
Preliminary, in our solution, we consider a set of service types and we impose that {\em each service $\sigma$ belongs to exactly one of this type, say $\Phi$}, and we denote it with $\sigma \mapsto \Phi$.
In our design, the questions included in a survey are mapped on specific constraints for services, which are expressed in the same form used for the definition of SecSLAs. 
At this point, one or more user answers to the survey questions can be associated with one or more constraints and, therefore, allow for the derivation of security labels, as illustrated above. 
In addition, questions are structured so that also the ``importance'' for the user of each constraint can be derived. This concept can be used to estimate, for each ordered space associated with the set of labels mapped to a property, the behavior of the distance function $d(\cdot,\cdot)$.
Practically speaking, in our solution, along with the order relationship $\succ^p$, we also define a ``cost'', in terms of a distance, when passing from one label to another in a chain of a label set.
The derived distance function can, hence, be used to build the weighted security lattice, modeling relationships and distances between classes, introduced before.

According to the reasoning above, the set of the answers to the survey questions implicitly corresponds to a {\em security class} for each type of service in the network. Indeed, a user could access more than one type of service in the network (e.g., video streaming, digital signature, etc.) that could have very different characteristics, hence, for each of them the user could express various security and privacy needs. For this reason, the user survey should take into account all the different offered services in such a way that a {\em security class} can be derived for each type of them.
Given a user survey $u$, we say that it satisfies a security class $c_i$ for a service type $\Phi$, denoted by $u \models^{\Phi} c_i$, if all the constraints derived by the answers in $u$ can be combined and mapped to the security labels of $c_i$.
More formally we can define the security class of the requirements of a user for a service type as:

\begin{definition}{\bf{(Security class of user requirements for service type)}}
    Given a user survey $u$, a service type $\Phi$, and a weighted security lattice $\mathcal{K}$ over $\mathcal{P}$, the security class of $u$ for $\Phi$, denoted as $\lambda(u^{\Phi})$, is the security class $c_i \in \mathcal{C}$ such that $u \models^{\Phi} c_i$, and $\nexists c_j \in \mathcal{C}$ with $c_j \succeq c_i$ and $u \models^{\Phi} c_j$. 
\end{definition}

In our system, we consider that the reference survey template is composed of the technical questions from which we derive the security constraints valid for a user regarding a service type, along with information on the importance conferred on each of these questions by the user.
This additional datum allows us to estimate how strong the requirements of a user are; that is how firmly the agent should match the security class of the user requirements when selecting a service provider during its life cycle.

\begin{figure*}[ht]
    \centering
    \includegraphics[width=0.65\textwidth]{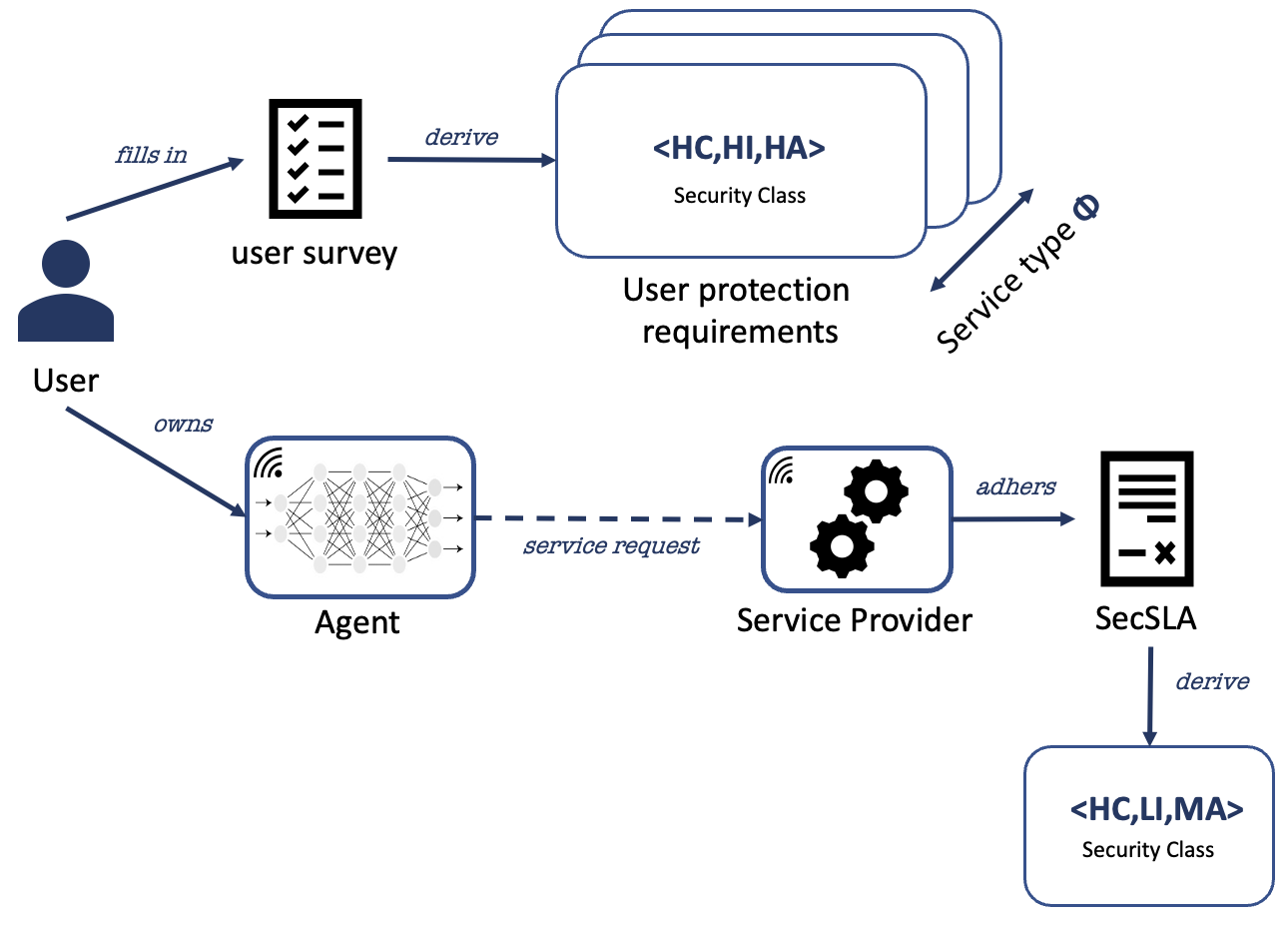}
    \caption{The general scheme of our framework. \label{fig:scheme}}
\end{figure*}

\noindent
Figure \ref{fig:scheme} shows how the different actors interact with each other to specify privacy and security requirements for the selection of suitable services (upper branch of this figure), and how services' characteristics (stated in a SecSLA) map onto security classes (lower branch). Note that, for the sake of presentation, the user protection requirements block contains only one security class. It means that in the example only one type of service is considered.

Finally, in our model, we also describe the possible security loss caused by a (partial) miss-match between the security requirements granted for a service by its provider and those specified on the user side. In particular, given a weighted security lattice $\mathcal{K}$, all the pairs of security class vertices are linked through weighted paths in $\mathcal{K}$. The sum of their weights represents the effective distance in terms of the corresponding security level. Therefore, the higher the traversal cost of the minimum path between two classes, the greater the difference between the security constraints provided by them. 
More formally, we give the following definition:

\begin{definition}{\bf{(Security Loss)}}
\label{def:secloss}
Given a weighted security lattice $\mathcal{K}$ over $\mathcal{P}$ and two security classes $\lambda(\sigma)$ and $\lambda(u^{\Phi})$, the former related to a service $\sigma \mapsto \Phi$ and the latter referring to the user security requirements on the same service type.
Let $w_{i,j} \in \mathcal{W}$ be the weight of a link in the lattice corresponding to the distance associated with two generic classes $\lambda_i$ and $\lambda_j$ of $\mathcal{C}$, the security loss is defined as the total cost of the minimum path $\rho(\sigma,u) = \{\langle \lambda_1, \lambda_2 \rangle,\dots,\langle \lambda_{n-1},\lambda_n)\rangle \ | \ \lambda_1 = \lambda(\sigma) \wedge \lambda_n=\lambda(u^{\Phi})\}$ linking $\lambda(\sigma)$ to $\lambda(u^{\Phi})$ in the weighted lattice $\mathcal{K}$:

$$\widehat{\xi_{\mathcal{K}}}(\sigma,u)=
\begin{cases}
    \sum_{(\lambda_i, \lambda_j) \in \rho(\sigma,u)} w_{i,j} & if \ \lambda(\sigma)<\lambda(u^{\Phi}) \\
    0 & otherwise
\end{cases}
$$
\end{definition}

\noindent

In the definition above, the security loss is estimated only if the security class of the target service $\sigma$ is lower than the one required by the user.
Of course, in the opposite case, such a loss value will be equal to zero.

Moreover, starting from this definition, we introduce the normalized security loss as follows:

\begin{definition}{\bf{(Normalized Security Loss)}}
\label{def:normsecloss}
Given a weighted security lattice $\mathcal{K}$ over $\mathcal{P}$, let $\rho_{max}$ be the length of the maximum minimum path between any pair of classes in $\mathcal{K}$.
Given two security classes $\lambda(\sigma)$ and $\lambda(u^{\Phi})$, the normalized security loss is:

\begin{equation}
\label{eq:nsecloss}
\xi_{\mathcal{K}}(\sigma,u)=
\begin{cases}
    \frac{\widehat{\xi_{\mathcal{K}}}(\sigma,u)}{|\rho_{max}|} & if \ \widehat{\xi_{\mathcal{K}}}(\sigma,u)\leq |\rho_{max}| \\
    1 & otherwise
\end{cases}
\end{equation}

\end{definition}

\noindent The normalized security loss refines Definition \ref{def:secloss} by considering the size of the security lattice adopted to describe security class relationships. Indeed, the higher the number of involved classes, the greater the size of the security lattice, and the higher the normalization factor that must be considered to obtain the relative security loss.
Definition \ref{def:normsecloss} provides a piece of important information to compute the rewards to train the behavior of an agent according to our proposal.
In the next sections, we will describe our DRL solution, in detail.

\section{A Deep Reinforcement Learning Solution} 
\label{sec:drl}

As stated in the previous sections, the objective of our framework is to train a local agent for IoT devices capable of making security-aware decisions according to the user needs (in terms of operations to be obtained through services) and requirements, the SecSLA of the services available in the environment.
In our scenario, each operation, a service user is interested in, is associated with an expiration time (i.e., the time after which the operation is no longer useful for the user).
We design our strategy like a Deep Reinforcement Learning (DLR, for short) problem focused on the designing of an agent capable of identifying the providers to interact with for obtaining the optimal services to complete the needed operations.

Like every DRL problem, to describe our solution, we start by defining its main components as follows:

\begin{itemize}
    \item {\em agent}: a software component for both learner and decision maker. In our scenario, we assume that each agent is associated with a user and, therefore, it is installed on the IoT devices of such a user. The agent is instructed to carry out tasks on behalf of its owner and is capable of making intelligent decisions through DRL. As such, an agent interacts with the environment through actions and receives rewards based on them.
    \item {\em environment}: a representation of the considered scenario surrounding IoT nodes from which an agent can learn the best strategies to carry out its tasks.
    \item  {\em state}: the condition of the agents, in terms of the values of its internal parameters, at a specific time-step in the environment. A new state is reached by the agent after performing an action on the environment and possibly getting a reward.
    \item {\em action}: an accomplishment pursued by the agent to receive a reward.
    \item {\em reward}: a numerical value that the agent receives on performing some action at some state(s) in the environment. 
\end{itemize}

\begin{figure}[ht]
    \centering
    \includegraphics[width=0.45\textwidth]{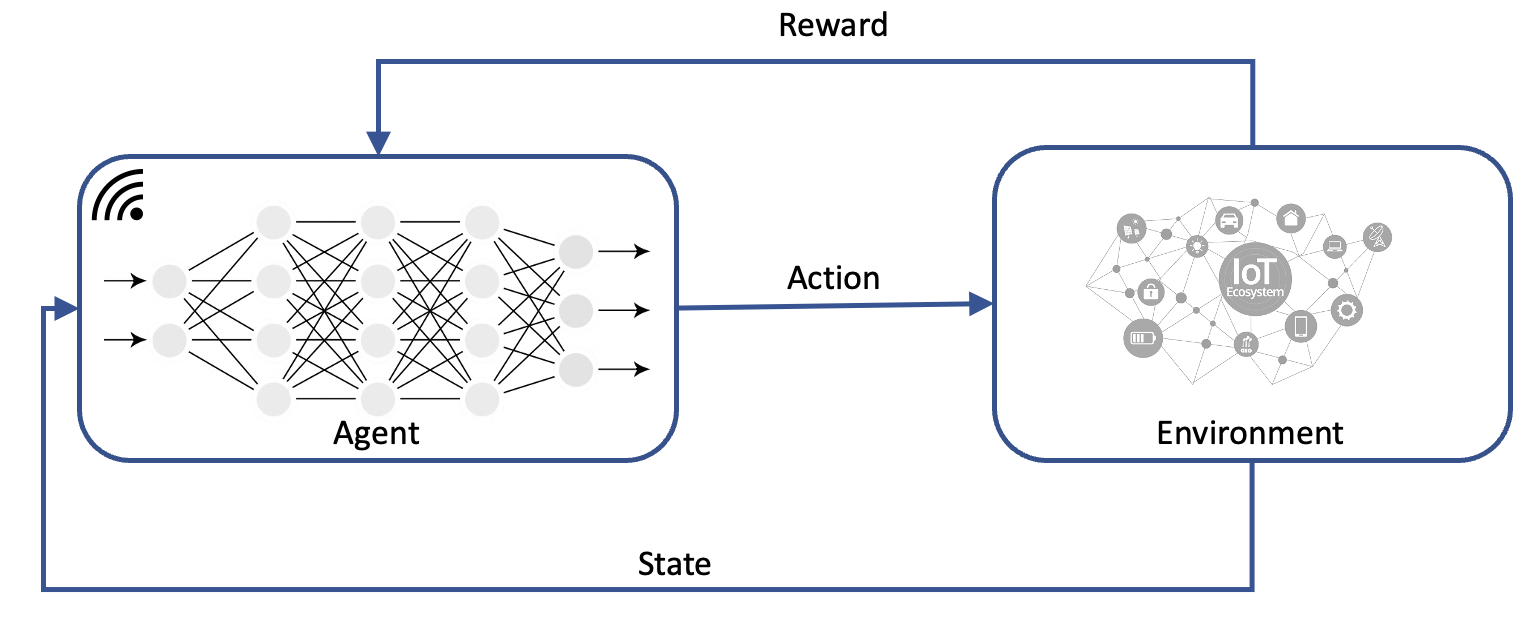}
    \caption{Representation of our Deep Reinforcement Learning scheme \label{fig:DRL_Scheme}}
\end{figure}

Reinforcement Learning is an AI experience-driven technique in which an autonomous agent makes an experience through a trial-and-error process aiming at improving its future choices.
In this situation, the problem to solve
is formulated as a discrete-time Markov stochastic control process defined through the tuple $(S, A_s, pr,r)$. In this tuple $S$ represents finite state space, $A_s$ is a finite action space for each state $s \in S$;  $pr$ is the state-transition probability from state $s$ to state $s' \in S$ taking an action $\alpha \in A$; and $r \in R$ is the immediate reward value obtained after an action $\alpha$ is performed. The agent's primary goal is to interact with the environment, at each time step $t = 0, 1, 2, \dots,$ to find the optimal policy $\pi^*$ and maximize the cumulative rewards in the long run. A policy $\pi$ is a function that represents the strategy used by the RL agent. It takes the state $s$ as input and returns an action $\alpha$ to be taken. The objective of the agent is to maximize the expected discounted return $R_t$ as follows:

 $$R_t = \sum_{i=0}^{\infty}\gamma^ir_{t+i+1}$$

\noindent
where $\gamma \in [0, 1)$ is the discount factor that determines how much importance is to be given to the immediate reward with respect to the future ones. The larger this parameter is, the more the estimated future rewards will be concerned.
Traditionally RL methods cannot find the optimal value functions or policy functions for all states within an acceptable delay. Thus, DRL
was developed to handle high-dimensional environments
based on the Deep Neural Network (DNN) value-function approximation. An example of a typical architecture for a DRL solution is shown in Figure \ref{fig:DRL_Scheme}.

In our solution, we leverage DQN that exploits Neural Networks to analyze inputs and derive an approximate
action-value $Q(s, \alpha|\theta)$ by minimizing the loss function $L(\theta)$ defined as follows:

$$L(\theta) = \mathbb{E}[(r+\gamma\max_{\alpha'}Q(s', \alpha'|\theta')-Q(s, \alpha|\theta))^2]$$

\noindent
where $\mathbb{E}[.]$ denotes the expectation function, $\theta$ and $\theta'$ represent the parameters of the predict and the (Temporal Difference - TD) target network, respectively. $Q(s, \alpha|\theta)$ is a state-action value function or Q-Function returning the value corresponding to a certain action $a$ in a state $s$ for the agent. The Q-learning in this method aims to directly
approximate the optimal action-value $Q^*(s, \alpha) \approx Q(s, \alpha|\theta)$. A DRL algorithm extracts features from the environment using DL while the RL agent optimizes itself by trial and error.
The goal of such an environment is to provide a representation of the considered scenario so that the agent can learn the best strategies to solve its tasks. 
With this in mind, the accurate definition and design of the actions and the associated rewards represent the key points for this environment to support the training of an agent capable of addressing the objectives introduced before, and, in particular, solving the trade-off between obtaining access to the needed services and the compliance with the security constraints.
The definition and suitable tuning of the reward function are critical to avoid common agent mistakes. Indeed, an unrefined function could lead the agent to either discard the security requirements to privilege the obtaining of a service or decline any service offered to minimize security concerns.
Of course, the definition of the reward functions is not the only important component of the design. 
The reward function must be integrated with an observation space that provides the agent with information about the status of the environment. Such information is crucial for the correct interpretation of the different scenarios the agent might be involved in, leading to the identification of the best strategies.
Practically speaking, in an RL scenario, the deep Q-network exploits the observation space as input to decide on the best action to be performed.
In our solution, we use a Neural Network composed of 2 hidden layers of 128 and 64 neurons, linked by a batch normalization, a dropout, and a ReLU activation layer. The input and output layers are modeled according to the observation space and the possible actions, respectively.
In the following sections, we will provide all the details of the proposed observation space, the possible actions, as well as the reward functions associated with them.

\subsection{Actions}
\label{sub:actions}

In this section, we describe all the possible actions an agent can perform along with the corresponding reward functions.
Rewards must be associated with conditions related to the considered environment, so that, by inspecting its status, the agent can identify the best actions to be carried out.
As stated above, the environment is, therefore, one of the main components of DRL solutions. 
From a technical point of view, an environment can be represented by its status, which must be interpreted by the agent.
Therefore, in the literature, DRL solutions are typically designed by focusing on the concept of observation space, that is the vectorial representation of the status of the considered environment.
Roughly speaking, it is composed of aspects that characterize both the current status of the agent and the status of the encountered entities.
Table \ref{tab:env} reports the elements characterizing the considered observation space.

\begin{table}[!ht]
\centering
\scriptsize
\begin{tabular}{||l|l||}
\hline
\bf{Element } & \bf{Description}\\
\hline \hline
$t$ & Current time step\\
$AS_t$ & The set of available service offers for the agent at time $t$ \\
$\mathcal{T}$ & The set of possible operations available from the services in\\
& the environment \\
$\mathcal{E}_u$  & The set of expiration times for the operations of $\mathcal{T}$ according\\
& to the user $u$ \\
$\vec{\tau_u}$ & The vectorial representation of the operations required to the\\
& agent along with their importance \\
$\vec{\tau_\sigma}$ & The one-hot encoding representation of the operation \\
& included in a service $\sigma$ \\
\hline
\end{tabular}
\caption{Elements characterizing the observation space.\label{tab:env}}
\end{table}

In our solution, during its interaction with the environment, an agent can meet providers exposing services of interest to the user. 
In such cases, the agent has to make a decision on whether to acquire the offered service or not based on how valuable the service is for its owner, the security aspects related to its acquisition, and the time deadline expressed by the user and after which such a service would be no longer useful.
An advanced feature in our strategy is that the agent can maintain a list of the providers (along with the offered services) it has met during its previous interactions with the environment. 
This list can be used by the agent to postpone the decision to acquire a specific service from a provider in an attempt to find better options in subsequent steps. 
Clearly, at each step, a possible action could be the extraction of a service provider from such a contact list to initiate the acquisition of a target service from it.

In summary, the agent can perform one of the following actions:

\begin{itemize}
    \item Acquiring the service offered by the currently met provider.
    \item Acquiring a target service from one of the providers in the list of contacts.
     \item Performing no action (i.e. refusing to interact with both any currently met service provider or any previously met one available in its contact list).
\end{itemize}

\subsection{Rewards}
\label{sub:rewards}

In this section, we define the reward functions used in our approach. 
As will be clear below, in our solution we include two separate strategies: the former targeting the case in which the agent accepts a service either from a current provider or from one belonging to its contact list ({\em Accept-Action}); the latter, instead, focusing on the case in which the agent does not accept any service provisioning in the current step ({\em Decline-Action}).

\subsubsection{Reward Function: Accept-Action}
\label{subsub:accept}

In our solution, the event in which an agent contacts a known service provider included in the list of its contacts shares the same outcome as the action of accepting the service provided by the IoT object it currently meets. Hence, we define a common reward function for both these cases, namely {\em Accept-Action}, including the following components: {\em(i)} the reward associated with the successful acquisition of a required service and the included operations; {\em(ii)} the reward associated with the inverse of the possible security loss (see Section \ref{sub:pla}) caused by the service fruition.

We start by defining the first contribution, which rewards the agent according to the operations it is capable of completing through the service acquired at a given step, weighted according to their importance for the user.
Moreover, it considers whether the operations included in such a service have been already completed before, thus resulting in a negative impact on the reward.
In this component also the expiration time (i.e., the deadline) plays a key role. Indeed, for each considered operation, the nearer its expiration time, the higher the priority it should have.

With that said, we can now define the different contributions in the formulation of this first reward component.
In particular, let $\mathcal{T}=\{\tau_1, \tau_2,\cdots,\tau_m\}$ be the set of the possible operations in the environment; a user can define the subset $\mathcal{T}_u \subset \mathcal{T}$ of such operations that should be addressed by the agent through the services available, along with their priority (or importance).
Therefore, we introduce the vectorial representation of $\mathcal{T}_u$ as a vector $\vec{\tau_u}$ of dimension $|\mathcal{T}|=m$. In this vector, the $i_{th}$ element is equal to $w_i$ if the users require the operation $\tau_i$ with importance $w_i \in (0,1]$, and $0$ if $\tau_i$ is not required. 
Additionally, as the agent proceeds in the service acquisition, the completed operations become an important input for the proper evaluation of the subsequent service offers. Indeed, such operations must be labeled as no longer required.
To encode this aspect, given a completed operation $\tau_i$, we enforce that the agent changes the corresponding value in $\vec{\tau_u}$ to zero.

At this point, given the set of service offers that the agent can evaluate at time $t$ (either from the list of contacts or from the current interaction), say $AS_t$, let $\sigma \mapsto \Phi$ be a service of $AS_t$, we define the one-hot encoding representation of the operations obtainable through $\sigma$ (as stated in its SLA) with a vector $\vec{\tau_{\sigma}}$ of dimension $|\mathcal{T}|=m$. In this vector, the $i_{th}$ element is equal to $1$ if $\sigma$ comprises the operation $\tau_i$, and $0$ otherwise. 

As stated before, each operation a user is interested in has an expiration time, after which such operation can be considered no longer required.
Our goal is to let the agent learn how to optimize the service acquisition to allow the completion of all the required operations before their expiration time.
For this reason, the reward is positively impacted by an early satisfaction with the required operations and, vice versa, reduced when approaching the operation expiration time.
Given the expiration time, namely, $\overline{t_i}$, of an operation $\tau_i$, let $t$ be time at the current step, to encode the reasoning above, we can formulate the reward decay based on the expiration time through a sigmoid function \cite{ezeafulukwe2018analytic}. In particular, we define the function $g(t,\vec{\tau_u}): \mathbb{R}^m \rightarrow \mathbb{R}^m$, such that an element of its output $g(t,\vec{\tau_u})_i$ is obtained as follows:

\begin{equation}
g(t,\vec{\tau_u})_i = \begin{cases} 
      (1+e^{t-\frac{\overline{t_i}}{2}})^{-1} & if \ t < \overline{t} \\
       0 &  otherwise \\
\end{cases}
\end{equation}

In this equation, the term $\frac{\overline{t_i}}{2}$ represents a safeguard interval during which the award decay is limited (i.e., the value for this contribution will be near $1$). Hence, during that period, the agent can evaluate the different options available to complete the required operation $\tau_i$. 
After this safeguard time, the reward will be exponentially impacted in a negative way; indeed, for $t=\frac{\overline{t_i}}{2}$ this contribution will be equal to $0.5$. 
In this way, the agent will attempt to complete $\tau_i$ as soon as possible and, in any case, before $t = \frac{\overline{t_i}}{2}$. Elsewhere, if $\tau_i$ has expired, the result of the $g$ function is $0$, thus resulting in no gain for the agent.

Hence, the overall reward obtained by the contributions of the first component can be formulated as follows:

\begin{equation}
    \mathrm{R}_O(\sigma, u, t)= \frac{\left( \vec{\tau_u} \odot g(t,\vec{\tau_u}) \right) \cdot \vec{\tau_\sigma}}{||\vec{\tau_u}||_1}
    \end{equation}

\noindent
where $||\vec{\tau_u}||_1$ is the Manhattan norm, while $\odot$ indicates the value of the Hadamard product of the two vectors.
As for the first term, because it consists of a Hadamard product and because the elements of both $\vec{\tau_u}$ and $g(t,\vec{\tau_u})$ are zero in case the corresponding operations are not required, already addressed, or expired, it allows to preserve only the contribution of the still required operations. Instead, the second term encodes the information about the subset of operations provided by the current service; therefore, the scalar product produces a score that is equal to the sum of the importance of the overlapping operations between the two terms. Such a score is, hence, normalized with the sum of the importance values for all the required operations.

\begin{example}
    Consider a scenario with six possible operations included in the offered services of the environment, namely $[$temperature, humidity, pressure, time, printing, connectivity$]$. Suppose that the operations required for the agent along with their importance are $[$temperature=0.5, humidity=0.8, time=0.7$]$. Moreover, consider that at a given time step $t$, the agent associated with $u$ is evaluating a service $\sigma \mapsto \Phi$ offered by a known provider, and the operations included by it are $[$time, connectivity$]$. The corresponding vectors are equals to $\vec{\tau_u}=[0.5,0.8,0,0.7,0,0]$ and $\vec{\tau_\sigma}=[0,0,0,1,0,1]$. Finally, suppose that the expiration time for all the required operations in $\tau_u$ is 300 seconds.
    Now, imagine the agent is evaluating such a service offered at the time $t=0$ seconds.
    The reward component $R_O(t)$ will be:

    $$R_0(\sigma, u, 0)=\frac{[0.5,0.8,0,0.7,0,0] \odot [1,1,1,1,1,1]}{0.5+0.8+0+0.7+0+0}\cdot$$
    $$\cdot\frac{[0,0,0,1,0,1]}{0.5+0.8+0+0.7+0+0} =\frac{0.7}{2}=0.35$$
    
    \noindent
    Instead, if the same evaluation should be performed at $t=180$ seconds, we have:
    
    $$R_0(\sigma, u, 150)=\frac{[0.5,0.8,0,0.7,0,0] \odot [0.5,0.5,0.5,0.5,0.5,0.5]}{0.5+0.8+0+0.7+0+0}\cdot $$
    $$\cdot\frac{[0,0,0,1,0,1]}{0.5+0.8+0+0.7+0+0} = \frac{0.35}{2}=0.18$$

\end{example}

We can now proceed by discussing the second contribution of the reward associated with the {\em Accept-Action}, (i.e., the one related to the security loss).
Indeed, in Section \ref{sub:pla} we defined a strategy to compute the cost, in terms of the distance of the security requirements, between the service security class and the user preferences.
Therefore, starting from Equation \ref{eq:nsecloss}, we can define the security loss contribution as follows:

\begin{equation}
    R_{SL}(\sigma, u)=1 - \xi_{\mathcal{K}}(\sigma,u)
\end{equation}

In this equation, the reward component for the agent is related to how much the agent will risk in terms of security if it uses the offered service.

Finally, by combining the two contributions above, the overall reward for the {\em Accept-Action} can be estimated as follows:

\begin{equation}
R_A(\sigma, u, t) = \frac{R_O(\sigma, u, t) + R_{SL}(\sigma, u)}{2}
\end{equation}

\begin{example_continue}
    Consider now the case in which $\sigma \models c_i$, i.e. it satisfies the security class $c_i$, with $c_i=[HC,HI,MA]$.
    Suppose that for the security requirements of the user $u$, represented by the agent, on the service type $\Phi$, holds the following: $u \models^{\Phi} c_j$, with $c_j=[HC, HI, HA]$.
    Moreover, let $w_{j,i}=1$ be the distance between the classes $c_i$ and $c_j$.
    Because $c_j \succeq c_i$, from Equation \ref{eq:nsecloss} we have that $|\rho(\sigma, u)|=1$, while $|\rho_{max}|=9$. Hence, the overall reward at the time step $t=0$ is:

    $$R_A(\sigma, u, 0)=\frac{R_0(\sigma, u, 0) + R_{SL}(\sigma, u)}{2} = \frac{0.35 + 0.89}{2} = 0.62$$
\end{example_continue}

\subsubsection{Reward Function: Decline-Action} 

The second possible action in our solution is the {\em Decline-Action}, according to which the agent does not accept any service provisioning during a step.
Indeed, in its interaction with the environment, the agent is supposed to learn the distribution of services, along with the different involved operations, to estimate, at each step, the probability of obtaining better offers for services in subsequent steps.
Therefore, a possible choice during a step is to discard both any current service offer and an offer from a provider in the contact list.
Of course, to allow the agent to select the {\em Decline-Action}, a non-zero reward should be obtained.
Besides, if no service is accepted during a step the security risk is zero, and, hence, the agent should be rewarded for not exposing its owner to possible threats.
Therefore, the only contribution available for this action is related to security protection.

To properly estimate the reward according to the reasoning above, we must consider that, at each step, the agent can evaluate service offers from both a currently met provider or a provider met during the previous steps and available in the list of contacts.
For this reason, given the set of services that the agent can evaluate during a time step $t$, say $AS_t$, the reward for the action {\em Decline-Action} can be defined as follows:

\begin{equation}
R_N(u,t)= \min_{\sigma \in AS_t} \xi_{\mathcal{K}}(\sigma,u)
\end{equation}

Intuitively, the above reward is estimated so that if any service available to the agent should produce a meaningful $R_O$ contribution for the {\em Accept-Action}, the advantage of opting for the {\em Decline-Action} should be negligible. 

\subsection{Contact List Management}

As explained above, the agent also maintains a contact list including information about the providers offering services including operations still required.
Clearly, a newly met provider is included in the list only in case the action chosen by the agent at a time step $t$ is the {\em Decline-Action}.

The strategy underlying the management of such a contact list works as follows.
In the case in which the list is not full, then the new provider is added without any additional checks.
Otherwise, a replacement policy is applied.
In our solution, this policy leverages again the contribution $R_O(\sigma, u, t)$ described in Section \ref{subsub:accept} and it estimates the potential reward obtainable by the operations offered in each service available from the providers in the contact list.

Given a currently met service provider offering the service $\Tilde{\sigma}$, let $AC$ be the of services available through the providers in the contact list.
The currently met service provider will be added to the contact list if:

$$R_O(\Tilde{\sigma}, u, t) > \min_{\sigma \in AC} R_O(\sigma, u, t)$$ 

\noindent
in such a case, the provider offering the service with the minimum $R_O$ value will be removed from the contact list and replaced by the newly met provider.

\subsection{Final States}
\label{sub:finalState}

A DRL solution requires the definition of the final state of the process.
In practice, a final state is a condition according to which the agent execution is ended.
In our case, we can identify the following final states:

\begin{itemize}
    \item All operations completed (Success).
    \item Security Requirement Violation (Failure).
    \item Time Expired (Failure).
\end{itemize}

The first final state concerns the case in which all the elements of $\vec{\tau_u}$ are set to zero.
This state implies a successful situation in which all the required operations have been completed by the agent.
The second final state, instead, focuses on the security requirements that the agent must meet.
In our solution, it is possible to estimate the extent of a security loss through Equation \ref{eq:nsecloss}. 
If the choices of the agent cause an excessive impact in terms of loss of security, then the execution should be stopped.
In our solution, we impose that the total security violation (i.e., the sum of all the security losses, $\xi_{\mathcal{K}}(\sigma,u)$, during the different time steps) must be lower than a control threshold $th_{\xi}$.
It is possible to set $th_{\xi}=1$, which implies that, overall, during the previous steps, the agent accumulated a total security violation equal to the maximum distance in the reference security lattice $\mathcal{K}$ (see Section \ref{sub:pla}).
Finally, the last final state focuses on the operation expiration times. Indeed, if the time elapsed before obtaining a required operation should exceed its configured expiration time, then a failure is identified and the agent execution must be stopped.

\section{Experiments and Results}
\label{sec:Experiments}

This section is devoted to the description of the experiments carried out to test the effectiveness of our proposal.
In particular, in the next sections, we will describe the testbeds along with the data used to simulate real-life agent's movements and meeting points, and we will report the obtained results along with their critical analysis.

\subsection{Dataset}
\label{sub:Dataset}

The reference dataset for our experiments has been obtained through a simulation software written in Python and generating (\emph{i}) the agent movements,  (\emph{ii}) the meeting points, and (\emph{iii}) the possible service providers the agent can interact within the environment.
To make our simulator as realistic as possible, we built it by exploiting two existing datasets available online. 
In particular, following the referring scenario described in Section \ref{sec:UseCase}, we used real-life movements in the city of New York to simulate how the agent interacts with the environment\footnote{The reference dataset for this step is available at  \url{www.kaggle.com/datasets/elemento/nyc-yellow-taxi-trip-data}.}.
To simulate the fact that, during its movements, an agent can meet service providers, we used information about Active Meeting Points (AMP, hereafter) in the same city.
To identify possible AMPs, we leveraged a dataset containing active establishment locations (e.g., bars and restaurants) in the city of New York\footnote{The reference dataset for this step is available at \url{data.cityofnewyork.us/Public-Safety/NYPD-Complaint-Map-Year-to-Date-/2fra-mtpn}.}.

\begin{figure}[ht]
    \centering
    \includegraphics[width=0.45\textwidth]{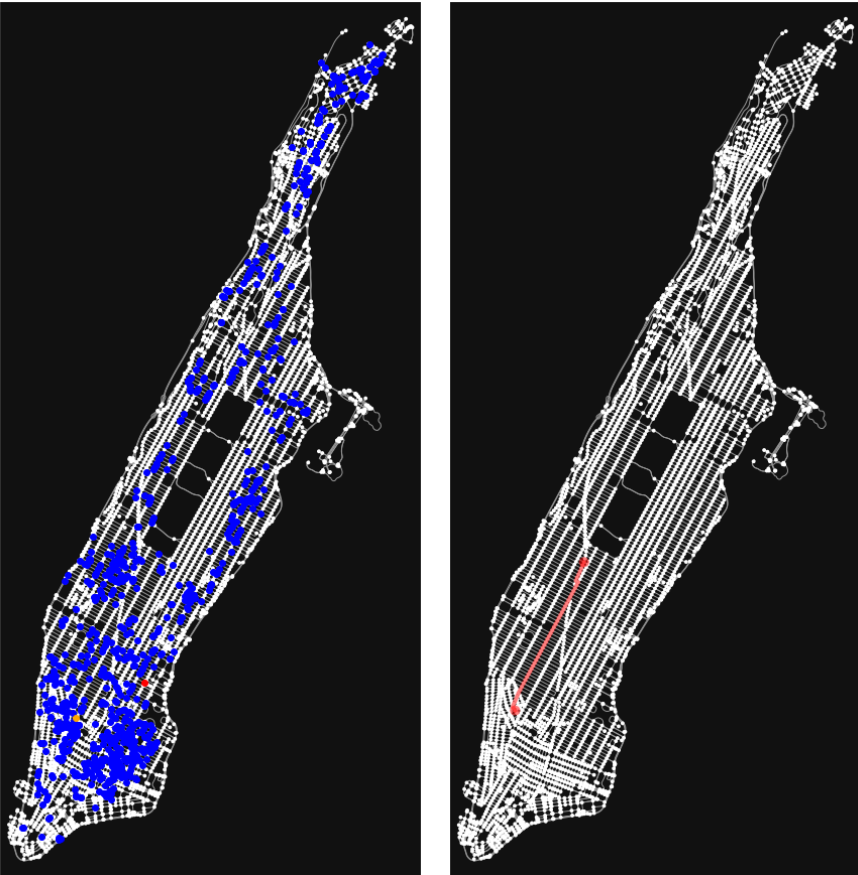}
    \caption{Two graphical representation of the simulated environment  \label{fig:ManhattanMap}}
\end{figure}

Figure \ref{fig:ManhattanMap} shows two graphical representations of our environment. In the map on the left, the blue points represent the places where the agent can meet a service provider, whereas the map on the right side represents an example of a complete path (shown with a red line) of an agent from a starting point to the ending point.

\begin{figure}[!ht]
    \centering
    \includegraphics[width=0.5\textwidth]{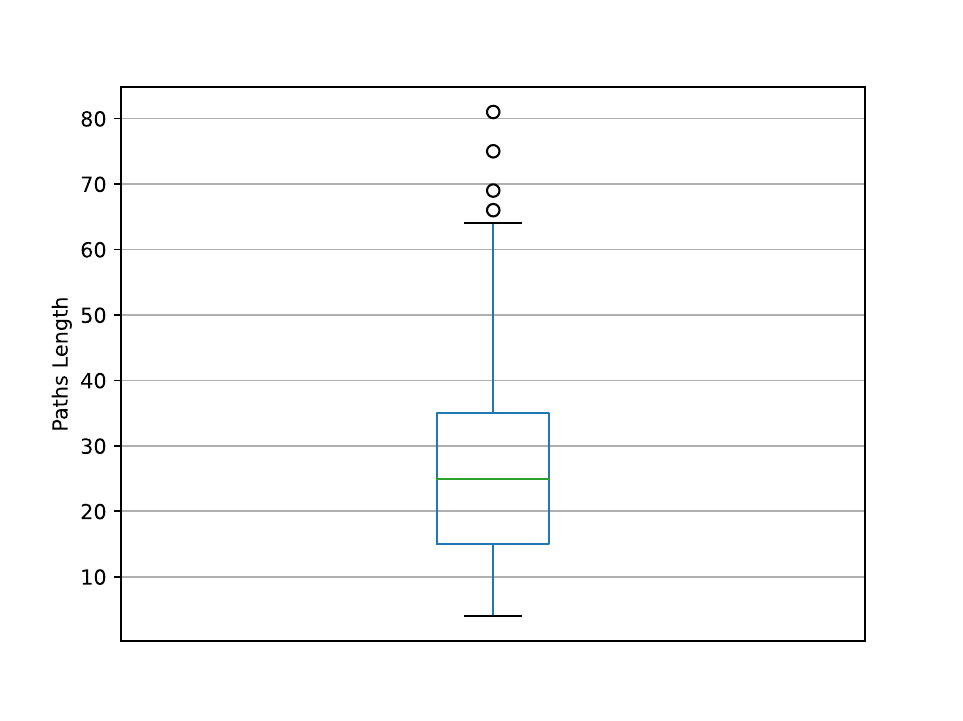}
    \caption{A box plot showing the distribution of the path length in our reference dataset \label{fig:DatasetLenghtsBoxplot}}
\end{figure}

By combining the two datasets above, we discarded the paths not traversing any AMP, thus obtaining a final dataset of $186$ paths of different lengths. The boxplot reported in Figure \ref{fig:DatasetLenghtsBoxplot} shows the distribution of the different path lengths in terms of steps.

Given the set of paths and the AMP locations, we simulated information about the services, the operations, and the security classes.
To do so, we proceeded as follows.
In our simulation software, we generated a random set of possible services in the environment and associated a set of possible operations with them. For each operation, a minimum security class is generated for the services including it.
At this point, during each training task, the simulator extracts a random vector $\vec{\tau_u}$ of required operations for the agent, among the ones available in the environment.
After that, for each service, offered by a service provider available in an AMP location, the software generates a service security class, by imposing the condition that it must be higher than or equal to the security classes generated for the included operation\footnote{In our simulator, we merged the concept of service and service provider, thus imposing that a provider grants exactly one service. Moreover, the condition on the minimum security class is imposed so that two services offering the same operation will both have a service security class higher than or equal to a common reference value.}.

As for the service association with AMPs (along with the involved operations), we considered two possible scenarios.
In the former, referred to as {\tt Scenario UDR}, such association is fully at random, and it is conceived to test the capability of the agent to learn the best decision strategies from a situation in which the services, and the corresponding operations, are uniformly distributed among the service providers available at the existing AMPs.
The latter, instead, in the following referred to as {\tt Scenario SKR}, is devoted to testing the capability of the agent to learn how to adjust its choices based on special configurations of the service offers and their security aspects.
In particular, the environment is configured to have skewed distributions of the services with variable security classes.
Such  ``extreme'' situations are designed to test the capability of the agent to learn how to optimally leverage its contact list.

In the next sections, we describe the experiments carried out to test the performance of our solution, along with the obtained results.

\subsection{Scenario UDR}
\label{sub:ExpUDR}

This section deals with the results of the experiments carried out on the {\tt Scenario UDR} introduced above.
Preliminarily, as classically done in the scientific literature, we adopt the term {\em episode} to identify a complete simulation, from the initialization, in which the agent is introduced in the environment with the list of required operations along with their priorities and deadlines, to the final stage in which the process is concluded because the agent has reached one of the final states described in Section \ref{sub:finalState}.
Of course, each episode generates important information that is used to train and improve the behavior of the agent in the subsequent ones.
With that said, we start by thoroughly examining the training process of the agent and by analyzing its behavior across $1,000$ consecutive training epochs, each composed of $200$ episodes.
For each episode, we inspect:

\begin{itemize}
    \item the $COP$ metric, that is the fraction of the completed operation;
    \item the $TLO$ metric, that is the total security loss caused by the exploitation of the different services, offering the required operations.
\end{itemize}

Formally, given the set of services $\{\sigma_1,\dots,\sigma_n\}$ exploited by the agent during an episode, the total security loss is computed as $TLO=\sum_{i=1}^{n} \xi_{\mathcal{K}}(\sigma_i,u)$.

\begin{figure*}
    \centering
    \subfigure[Training process for the agent without contact list]{
        \includegraphics[width=0.45\linewidth]{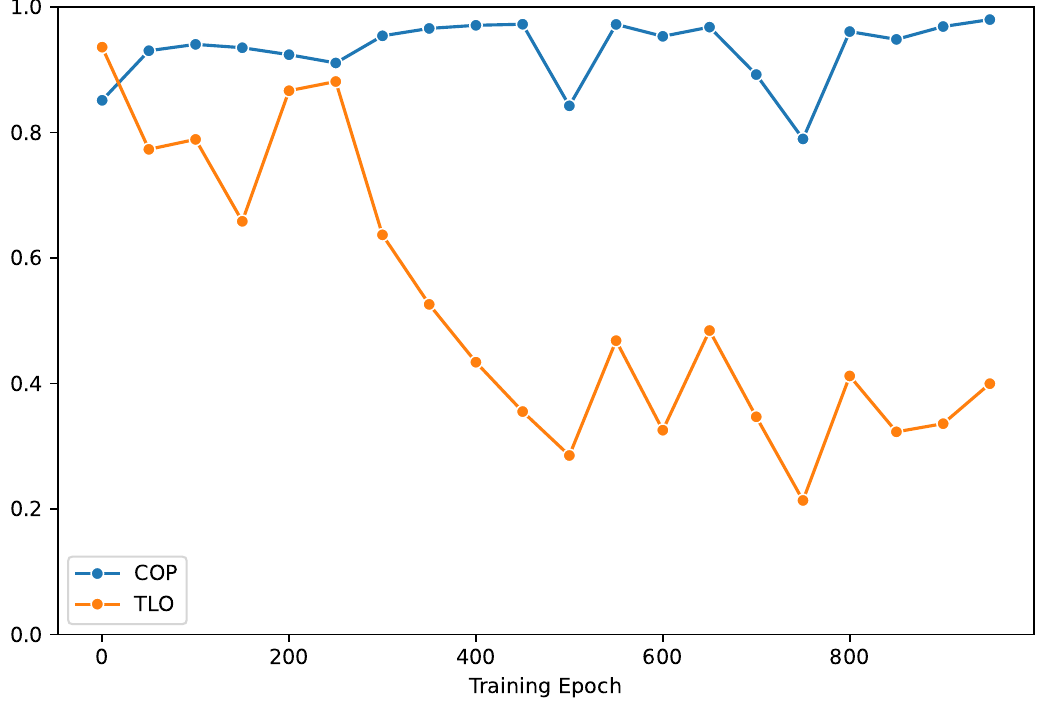}
        \label{fig:Objectives_privacy_graphA}
    }
    \hfill
    \subfigure[Training process for the agent with contact list]{
        \includegraphics[width=0.45\linewidth]{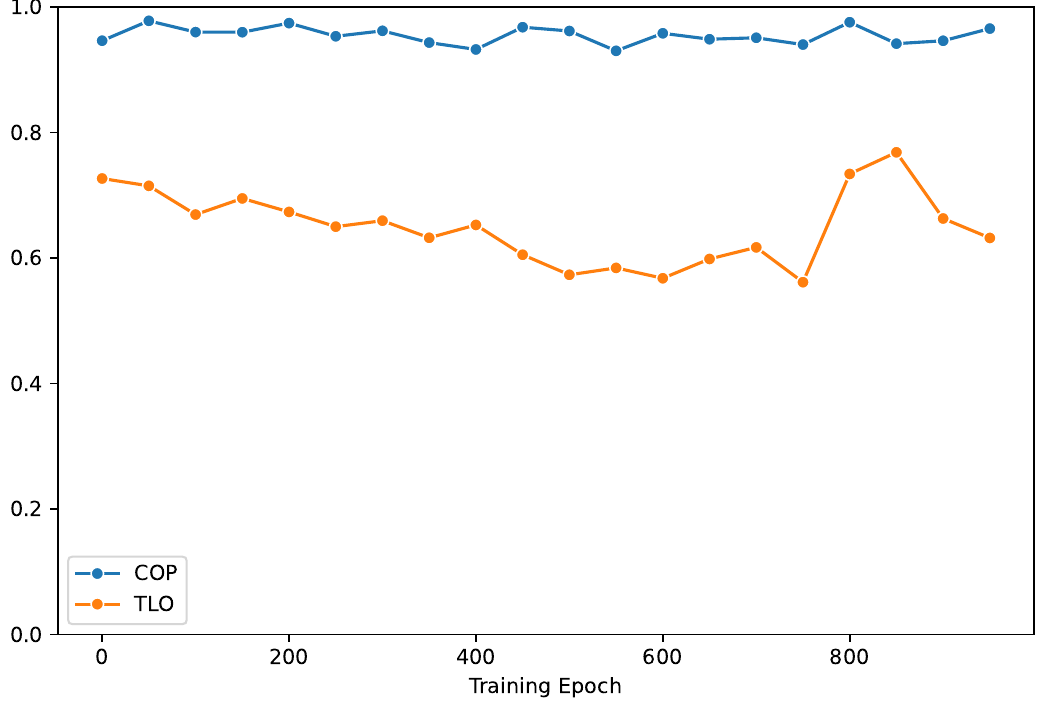}
        \label{fig:Objectives_privacy_graphB}
    }
    \caption{Training process for the agent with and without contact list}\label{fig:PlotObjectivesPrivacy}
\end{figure*}

The obtained results are reported in Figure \ref{fig:PlotObjectivesPrivacy}.
Moreover, to understand the importance of the contact list in the performance of the agent, we repeated the analysis with two configurations: {\em (i)} contact-list size equal to zero (Figure \ref{fig:Objectives_privacy_graphA}), and {\em (ii)} contact-list size equal to 3 (Figure \ref{fig:Objectives_privacy_graphB})\footnote{A thorough evaluation of the role of the contact-list size is carried out in the following experiments.}.

By analyzing this figure, we can see that, in the case without contact-list, the performance measured through the $COP$ metric oscillates and, eventually, after the first $800$ training epochs, the agent learns how to complete on average $98\%$ of the required operations ($COP$=$0.98$) across the $200$ validation episodes. This is achieved with a very low cost in terms of $TLO$, which is lower than $0.4$ on average.
By contrast, in the configuration with a non-empty contact list, the agent reaches very high values (about $0.98$) of the $COP$ metric right during the first episodes; this high value remains stable across all the analyzed episodes. However, such a positive result comes with an additional cost in terms of $TLO$, which remains always higher than $0.65$.

\begin{figure}[!ht]
    \centering
    \includegraphics[width=0.45\textwidth]{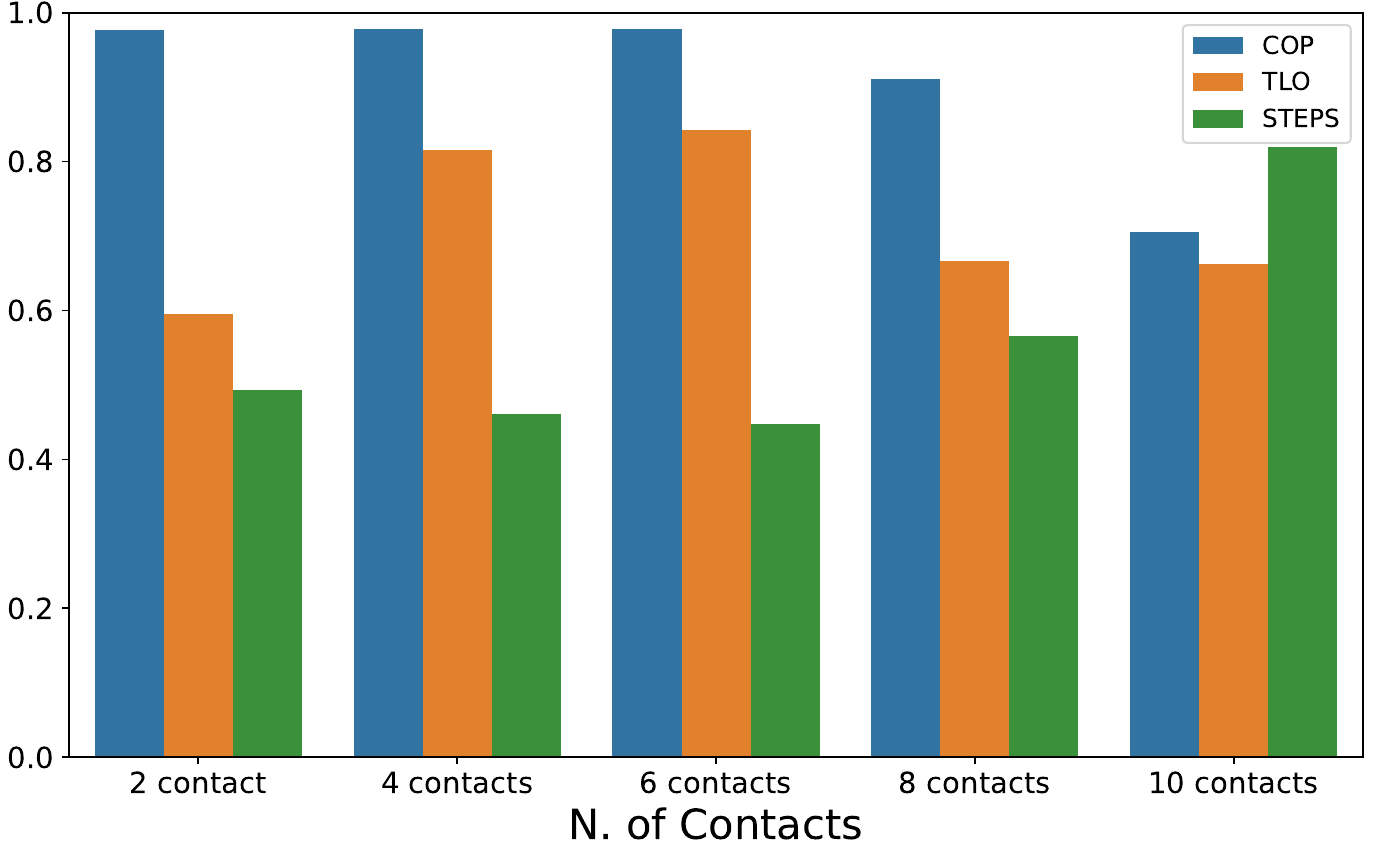}
    \caption{Performance impact for different sizes of the contact list \label{fig:PlotObjectivesPrivacyStepsContact}}
\end{figure}

To deep dive into the role of the contact list in our solution, once again in the first scenario, we also measured the $COP$ and $TLO$ metrics in several configurations characterized by a different size of the agent contact list.
In particular, we focused on 5 configurations in which the contact list size assumes the following values: $2$, $4$, $6$, $8$, and $10$.
Moreover, to better understand the agent behavior we recorded the average fraction of AMPs in the paths, namely $STEPS$, that the agent needs to exploit to maximize the number of required operations completed during an episode.
The obtained results are reported in Figure \ref{fig:PlotObjectivesPrivacyStepsContact}.

From the analysis of this figure and based on the previous results, we can conclude that a contact list with a small size helps the agent to quickly converge towards optimal $COP$ values at the price of a slightly higher $TLO$. However, at least in a random scenario in which services and operations are uniformly distributed and available in the paths, the contact list does not help the agent reach higher performance in terms of $COP$.
As for the average fraction of AMPs visited by the agent, this value grows with the size of the contact list. This could be explained by considering that the possibility of reaching back a previously met service provider makes the agent more inclined to postpone the decision of completing an operation through an offered service with the attempt of seeking a better opportunity.

\subsection{Scenario SKR}

To thoroughly investigate the performance of our solution, we proceeded by focusing on the second scenario described above, namely {\tt Scenario SKR}.
In a situation in which the distribution of required operations is non-uniform, the conditional probability of obtaining the required operation in a future step, given that a service provider is already offering such an opportunity in the current step, becomes extremely low. Therefore, intuitively, the role of the contact list may play a more crucial role in such a case.

To verify this intuition, we repeated the same experiments described above for this second scenario and measured both the average $COP$ and $TLO$ metrics.
In Table \ref{table:ResultsDifferentScenarios}, we report the obtained results for both {\tt Scenario UDR} and {\tt Scenario SKR} obtained when the agent holds a contact list and when it does not.

\begin{table}[!ht]
\centering
\begin{tabular}{|c|cc|cc|}
\hline
\multirow{2}{*}{} & \multicolumn{2}{c|}{No Contacts List}   & \multicolumn{2}{c|}{Contact List}  \\ 
\cline{2-5} 
Scenario & \multicolumn{1}{c|}{$COP$} & $TLO$ & \multicolumn{1}{c|}{$COP$}     & $TLO$ \\
\hline
{\tt UDR} & \multicolumn{1}{c|}{$98\%$} & $39\%$& \multicolumn{1}{c|}{$98\%$} & $65\%$ \\ 
\hline
{\tt SKR} & \multicolumn{1}{c|}{78\%} & $35\%$ & \multicolumn{1}{c|}{$86\%$} & $63\%$ \\
\hline
\end{tabular}
\caption{Fraction of completed operations
($COP$) and total security loss ($TLO$) for {\tt Scenario UDR} and {\tt Scenario SKR} with and without contact list
\label{table:ResultsDifferentScenarios}}
\end{table}

By analyzing this table, we can see that the impact of the availability of the contact list is critical for the {\tt Scenario SKR} to allow the agent to obtain a $COP$ value equal to $0.86$, on average.
The performance drop can be justified by the intrinsic higher complexity of this second scenario, which prevents the agent from completing more than $78\%$ of required operations without leveraging a support contact list.
Overall, we can conclude that, in complex scenarios, the possibility of storing previously met service providers in a contact list can help the agent counteract the expected performance drop and maintain a high capability of obtaining required operations.

\subsection{Multiple Operations}

As explained above in this paper, operations are obtained by the agent by accessing services available in the environment.
However, typically, a service consists of a set of included operations which, in turn, can be useful or not for the agent.
Of course, one of the objectives of our solution is to train the agent in such a way as to avoid accepting service proposals which, although useful to complete a required operation, include an excessive set of unneeded operations.
So the agent has to learn how to maximize the set of completed operations through service provisioning by limiting the ratio of non-required operations included in them.
Such a learning objective is encoded in the $\mathrm{R}_O(\sigma, u, t)$ component of the agent reward function described in Section \ref{sub:actions}. 
To test the performance of our solution in this sense, we focused, once again, on the {\tt Scenario UDR} and performed another experiment in which we also measured the fraction of unneeded operations, named $UNNO$, included in the services accepted by the agent to complete all the operations specified in $\vec{\tau_u}$.
Also in this case, we considered two configurations for our agent, i.e., with and without the possibility of using a contact list.

The results of this experiment are reported in Table \ref{table:ResultsANWOG}.

\begin{table}[!ht]
\centering
\begin{tabular}{|ccc|ccc|}
\hline
\multicolumn{3}{|c|}{No Contacts List}   & \multicolumn{3}{c|}{Contact List}  \\ \cline{1-6} 
                  \multicolumn{1}{|c|}{$COP$}       & \multicolumn{1}{c|}{$TLO$}   & $UNNO$       & \multicolumn{1}{c|}{$COP$}       & \multicolumn{1}{c|}{$TLO$}     & $UNNO$   \\ \hline
     \multicolumn{1}{|c|}{0.98} & \multicolumn{1}{c|}{0.39} & 0.30 & \multicolumn{1}{c|}{0.98} & \multicolumn{1}{c|}{0.65} & 0.14\\ \hline
    
\hline
\end{tabular}
\caption{Analysis of the fraction of unneeded operations ($UNNO$) with and without contact list in {\tt Scenario UDR} \label{table:ResultsANWOG}}
\end{table}

As seen above, for the {\tt Scenario UDR} the availability of the contact list does not help the agent to achieve higher $COP$ values.
However, as for the newly considered metric ($UNNO$), we can see that the agent with the contact list makes a better selection of the services to exploit, halving the fraction of included unwanted operations.
This result is, once again, obtained at the cost of a higher $TLO$, which is, however, always lower than the target final-state threshold $th_{\xi}$ (see Section \ref{sub:finalState}).
In conclusion, we can observe how the availability of the contact list helps the agent move in more complex scenarios and make more refined decisions, thus justifying its inclusion in our solution.

\subsection{Comparison with basic approaches}

To demonstrate the effectiveness of our proposal, in this section, we compare its performance against two basic approaches:

\begin{itemize}
    \item A naive ``Always Accept'' policy according to which the agent will accept any service offered until it reaches one of the final states of Section \ref{sub:finalState}.
    \item A ``Random Action'' policy, i.e., the agent will accept or refuse a currently offered service, at random.
\end{itemize}

One of the intended objectives of this experiment is also to validate the quality of the generated scenarios by showing that with basic and trivial approaches, the agent is not capable of solving them.

Therefore, in this experiment, we considered the same configurations analyzed in the previous experiments for both the simpler {\tt Scenario UDR} and the {\tt Scenario SKR}. Hence, we assessed the performance of both the basic approaches described above and the performance of our solution with and without the availability of a contact list.
The obtained results are reported in Table \ref{table:ResultsDifferentScenarios2}.

\begin{table*}[!ht]
\centering
\begin{tabular}{|c|cc|cc|cc|cc|}
\hline
\multirow{2}{*}{} & \multicolumn{2}{c|}{Always Accept}                             & \multicolumn{2}{c|}{Random Action}                            & \multicolumn{2}{c|}{No Contacts List}   & \multicolumn{2}{c|}{Contact List}  \\ \cline{2-9} 
                  Scenario & \multicolumn{1}{c|}{$COP$}       & $TLO$         & \multicolumn{1}{c|}{$COP$}       & $TLO$          & \multicolumn{1}{c|}{$COP$}       & $TLO$ & \multicolumn{1}{c|}{$COP$}       & $TLO$         \\ \hline
    {\tt UDR} & \multicolumn{1}{c|}{0.70} & \color{red}{\bf{1.19}} & \multicolumn{1}{c|}{0.60} & \color{red}{\bf{1.05}} & \multicolumn{1}{c|}{0.98} & 0.39 & \multicolumn{1}{c|}{0.98} & 0.65 \\ \hline
    {\tt SKR} & \multicolumn{1}{c|}{0.53} & \color{red}{\bf{1.29}} & \multicolumn{1}{c|}{0.51} & \color{red}{\bf{1.08}} & \multicolumn{1}{c|}{0.78} & 0.35 & \multicolumn{1}{c|}{0.86} & 0.63 \\
\hline
\end{tabular}
\caption{Performance comparison with basic approaches for both {\tt Scenario UDR} and {\tt Scenario SKR} \label{table:ResultsDifferentScenarios2}}
\end{table*}

From the analysis of this table, we can see that both the ``Always Accept'' and ``Random Action'' policies lead the agent to exceed the $th_{\xi}$ threshold for the total security loss (see red values in the table), thus reaching one of the {\em failure} final states of Section \ref{sub:finalState}.
In any case, even before the failure, the maximum reached COP for the simple {\tt Scenario UDR} does not exceed $0.7$, on average, for both approaches.
This result is even worse for the more complex {\tt Scenario SKR}, for which the two approaches always have a COP lower than $0.54$.

\subsection{Running Time}

As a final experiment, to demonstrate the effectiveness of our solution in the considered IoT realm, we measured the inference time of our reinforcement solution on different hardware settings.
To stick with the most common configurations in IoT, we did not consider devices possibly equipped with GPUs and, therefore, we only focused on the CPU inference time on different devices.
In particular, we considered three different heterogeneous machines, namely: {\em (i)} a Desktop PC equipped with an octa-core CPU (i.e., AMD Ryzen 7 5800X), {\em (ii)} a Raspberry Pi 4 with a Quad-core Cortex-A72 (ARM v8), and a very basic device equipped with a Single-Core ARM processor. The obtained measurements are visible in Table \ref{tab:hwInfTime}.

\begin{table}[]
\centering
\begin{tabular}{|c|c|}
\hline
Hardware        & Inference Time (ms) \\ \hline
Desktop CPU     & 0.5          \\ \hline
Raspberry Pi 4  & 13           \\ \hline
Single-Core ARM & 881          \\ \hline
\end{tabular}
\caption{Inference execution time on different hardware configurations \label{tab:hwInfTime}}
\end{table}

The results reported in this table allow us to conclude that our approach can be deployed also on very basic and simple devices, which are commonly available in an IoT setting.
Even the less powerful device can use our model to decide on a service offer in less than 1 second.

A final comment should be made for the contact list, which requires the availability of some sort of memory on the device. Of course, this could be critical for extremely basic and legacy IoT devices (think, for instance, of a smart thermometer). However, as shown in Section \ref{sub:ExpUDR}, a very limited number of contacts (up to five) is needed to improve the performance of our solution for complex scenarios (e.g., {\tt Scenario SKR}).
Moreover, in our proposal, the devices in which the agent is deployed must be capable of supporting their owners in daily life activities and, therefore, they can be assumed to be modern IoT smart objects, which are typically equipped with sufficiently powerful hardware and memory availability.

\section{Conclusion}
\label{sec:Conclusion}
In the IoT scenario in which smart things are more and more autonomous and pervasive in everyday life, preserving user privacy, ensuring the security of users and devices, and guaranteeing the availability of the services offered by this new ecosystem is becoming a major concern.
In this paper, we start from the above consideration to design a complete framework allowing a user of the system to express her/his need for privacy and security and choose IoT services accordingly. Specifically, the environment the user interacts with is populated with smart things providing the most disparate services and applications. As usual in this context, a Service Level Agreement (SLA) is used to regulate the interaction between a service provider and a client. In our case, we leverage two particular SLA declinations known as Security and Privacy SLA (SecSLA and PLA, respectively) to outline the level of security and privacy a smart object can provide. Every user of the network is equipped with an agent that according to the expressed user security and privacy requirement, the operations needed along with their priority, and a certain time deadline can select the most appropriate providers to satisfy the user's need. To achieve this objective our framework is based on a Deep Reinforcement Learning algorithm that allows an agent to learn from the environment some sub-optimal policies and make real-time decisions.
An extensive experimental campaign is then presented to evaluate the performance of our DRL-based approach.

The research directions explored in this paper can be regarded as a foundation, as we intend to continue our efforts aligned with this research in the future. For instance, we plan to expand our framework by designing a model and the corresponding language for the representation of protection requirements, encompassing both user and application aspects. Moreover, another interesting direction is to work on the definition and inclusion of other constraints to optimize not only security and privacy requirements but also the workload placement in the IoT network.

\section*{Acknowledgments}
\begin{figure}[ht]
    \includegraphics[width=3.6cm]{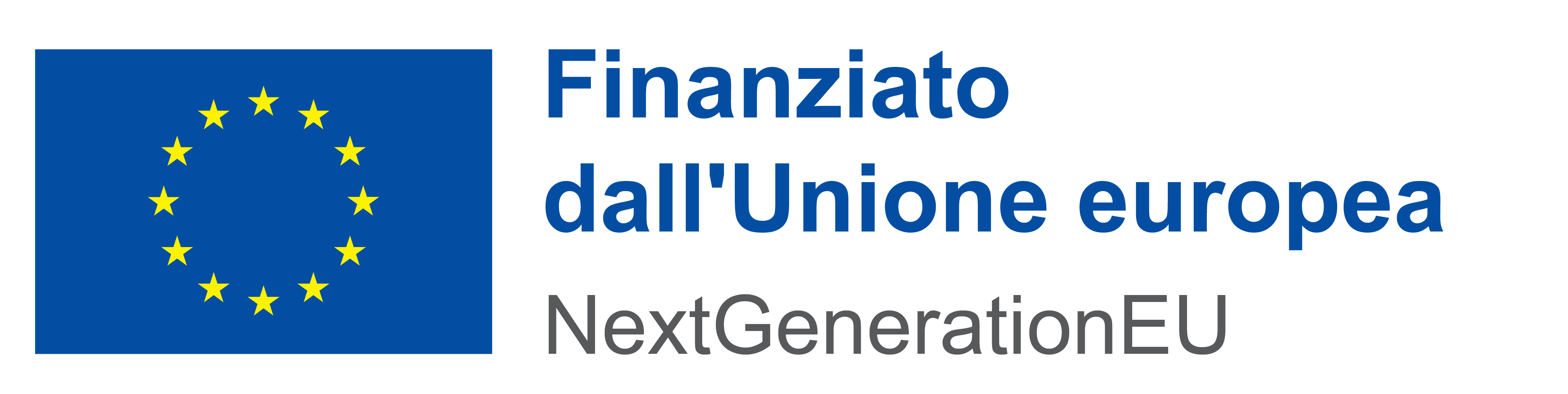}
\end{figure}

This work was supported in part by the SERICS project (grant number PE00000014) under the NRRP MUR program funded by the EU-NGEU, and by the PRIN 2022 Project ``HOMEY: a Human-centric IoE-based Framework for Supporting the Transition Towards Industry 5.0'' (code: 2022NX7WKE, CUP: F53D23004340006) funded by the European Union - Next Generation EU.



\begin{thebibliography}{10}

\bibitem{al2020blockchain}
Ismaeel Al~Ridhawi, Moayad Aloqaily, Azzedine Boukerche, and Yaser Jaraweh.
\newblock A blockchain-based decentralized composition solution for iot
  services.
\newblock In {\em Icc 2020-2020 ieee international conference on communications
  (icc)}, pages 1--6. IEEE, 2020.

\bibitem{arazzi2023fully}
Marco Arazzi, Serena Nicolazzo, and Antonino Nocera.
\newblock A fully privacy-preserving solution for anomaly detection in iot
  using federated learning and homomorphic encryption.
\newblock {\em Information Systems Frontiers}, pages 1--24, 2023.

\bibitem{arazzi2024novel}
Marco Arazzi, Serena Nicolazzo, and Antonino Nocera.
\newblock A novel iot trust model leveraging fully distributed behavioral
  fingerprinting and secure delegation.
\newblock {\em Pervasive and Mobile Computing}, page 101889, 2024.

\bibitem{barth2017privacy}
Susanne Barth and Menno~DT De~Jong.
\newblock The privacy paradox--investigating discrepancies between expressed
  privacy concerns and actual online behavior--a systematic literature review.
\newblock {\em Telematics and informatics}, 34(7):1038--1058, 2017.

\bibitem{barth2019putting}
Susanne Barth, Menno~DT de~Jong, Marianne Junger, Pieter~H Hartel, and Janina~C
  Roppelt.
\newblock Putting the privacy paradox to the test: Online privacy and security
  behaviors among users with technical knowledge, privacy awareness, and
  financial resources.
\newblock {\em Telematics and informatics}, 41:55--69, 2019.

\bibitem{bianco2008service}
Philip Bianco, Grace~A Lewis, and Paulo Merson.
\newblock {\em Service level agreements in service-oriented architecture
  environments}.
\newblock Carnegie Mellon University, Software Engineering Institute, 2008.

\bibitem{casola2018security}
Valentina Casola, Alessandra De~Benedictis, Massimiliano Rak, and Umberto
  Villano.
\newblock A security metric catalogue for cloud applications.
\newblock In {\em Complex, Intelligent, and Software Intensive Systems:
  Proceedings of the 11th International Conference on Complex, Intelligent, and
  Software Intensive Systems (CISIS-2017)}, pages 854--863. Springer, 2018.

\bibitem{chen2021deep}
Wuhui Chen, Xiaoyu Qiu, Ting Cai, Hong-Ning Dai, Zibin Zheng, and Yan Zhang.
\newblock Deep reinforcement learning for internet of things: A comprehensive
  survey.
\newblock {\em IEEE Communications Surveys \& Tutorials}, 23(3):1659--1692,
  2021.

\bibitem{di2017supporting}
Sabrina De~Capitani Di~Vimercati, Sara Foresti, Giovanni Livraga, Vincenzo
  Piuri, and Pierangela Samarati.
\newblock Supporting user requirements and preferences in cloud plan selection.
\newblock {\em IEEE Transactions on Services Computing}, 14(1):274--285, 2017.

\bibitem{di2019security}
Sabrina De~Capitani di~Vimercati, Sara Foresti, Giovanni Livraga, Vincenzo
  Piuri, and Pierangela Samarati.
\newblock Security-aware data allocation in multicloud scenarios.
\newblock {\em IEEE Transactions on Dependable and Secure Computing},
  18(5):2456--2468, 2019.

\bibitem{di2016supporting}
Sabrina De~Capitani di~Vimercati, Giovanni Livraga, Vincenzo Piuri, Pierangela
  Samarati, and Gerson~A Soares.
\newblock Supporting application requirements in cloud-based iot information
  processing.
\newblock In {\em IoTBD}, pages 65--72, 2016.

\bibitem{ezeafulukwe2018analytic}
Uzoamaka~A Ezeafulukwe, Maslina Darus, and O~Fadipe-Joseph.
\newblock On analytic properties of a sigmoid function.
\newblock {\em Int. Journal of Mathematics and Computer Science},
  13(2):171--178, 2018.

\bibitem{ferretti2021h2o}
Marco Ferretti, Serena Nicolazzo, and Antonino Nocera.
\newblock H2O: secure interactions in iot via behavioral fingerprinting.
\newblock {\em Future Internet}, 13(5):117, 2021.

\bibitem{frikha2021reinforcement}
Mohamed~Said Frikha, Sonia~Mettali Gammar, Abdelkader Lahmadi, and Laurent
  Andrey.
\newblock Reinforcement and deep reinforcement learning for wireless internet
  of things: A survey.
\newblock {\em Computer Communications}, 178:98--113, 2021.

\bibitem{fujii2018consideration}
Toru Fujii, TianBao Guo, and Akira Kamoshida.
\newblock A consideration of service strategy of japanese electric
  manufacturers to realize super smart society (society 5.0).
\newblock In {\em Knowledge Management in Organizations: 13th International
  Conference, KMO 2018, {\v{Z}}ilina, Slovakia, August 6--10, 2018, Proceedings
  13}, pages 634--645. Springer, 2018.

\bibitem{nist2013}
NIST Interagency~Report (IR).
\newblock Glossary of key information security terms.
\newblock \url{http://nvlpubs.nist.gov/nistpubs/ir/2013/NIST.IR.7298r2.pdf},
  2013.

\bibitem{kearney2010sla}
Keven~T Kearney, Francesco Torelli, and Constantinos Kotsokalis.
\newblock Sla$\star$: an abstract syntax for service level agreements.
\newblock In {\em 2010 11th IEEE/ACM International Conference on Grid
  Computing}, pages 217--224. IEEE, 2010.

\bibitem{lei2020deep}
Lei Lei, Yue Tan, Kan Zheng, Shiwen Liu, Kuan Zhang, and Xuemin Shen.
\newblock Deep reinforcement learning for autonomous internet of things: Model,
  applications and challenges.
\newblock {\em IEEE Communications Surveys \& Tutorials}, 22(3):1722--1760,
  2020.

\bibitem{liang2020deep}
Wei Liang, Weihong Huang, Jing Long, Ke~Zhang, Kuan-Ching Li, and Dafang Zhang.
\newblock Deep reinforcement learning for resource protection and real-time
  detection in iot environment.
\newblock {\em IEEE Internet of Things Journal}, 7(7):6392--6401, 2020.

\bibitem{liu2012sla}
Hui Liu, Fenglin Bu, and Hongming Cai.
\newblock Sla-based service composition model with semantic support.
\newblock In {\em 2012 IEEE Asia-Pacific Services Computing Conference}, pages
  374--379. IEEE, 2012.

\bibitem{liu2019deep}
Qingzhi Liu, Long Cheng, Tanir Ozcelebi, John Murphy, and Johan Lukkien.
\newblock Deep reinforcement learning for iot network dynamic clustering in
  edge computing.
\newblock In {\em 2019 19th IEEE/ACM international symposium on cluster, Cloud
  and Grid Computing (CCGRID)}, pages 600--603. IEEE, 2019.

\bibitem{ludwig2003web}
Heiko Ludwig, Alexander Keller, Asit Dan, Richard~P King, and Richard Franck.
\newblock Web service level agreement (wsla) language specification.
\newblock {\em Ibm corporation}, pages 815--824, 2003.

\bibitem{maddikunta2022industry}
Praveen Kumar~Reddy Maddikunta, Quoc-Viet Pham, B~Prabadevi, Natarajan Deepa,
  Kapal Dev, Thippa~Reddy Gadekallu, Rukhsana Ruby, and Madhusanka Liyanage.
\newblock Industry 5.0: A survey on enabling technologies and potential
  applications.
\newblock {\em Journal of Industrial Information Integration}, 26:100257, 2022.

\bibitem{mnih2013playing}
Volodymyr Mnih, Koray Kavukcuoglu, David Silver, Alex Graves, Ioannis
  Antonoglou, Daan Wierstra, and Martin Riedmiller.
\newblock Playing atari with deep reinforcement learning.
\newblock {\em arXiv preprint arXiv:1312.5602}, 2013.

\bibitem{mohammed2020ubipriseq}
Thaha Mohammed, Aiiad Albeshri, Iyad Katib, and Rashid Mehmood.
\newblock Ubipriseq‚Äîdeep reinforcement learning to manage privacy, security,
  energy, and qos in 5g iot hetnets.
\newblock {\em Applied Sciences}, 10(20):7120, 2020.

\bibitem{nguyen2020deep}
Thanh~Thi Nguyen, Ngoc~Duy Nguyen, and Saeid Nahavandi.
\newblock Deep reinforcement learning for multiagent systems: A review of
  challenges, solutions, and applications.
\newblock {\em IEEE transactions on cybernetics}, 50(9):3826--3839, 2020.

\bibitem{paschke2005rbsla}
Adrian Paschke.
\newblock Rbsla a declarative rule-based service level agreement language based
  on ruleml.
\newblock In {\em International Conference on Computational Intelligence for
  Modelling, Control and Automation and International Conference on Intelligent
  Agents, Web Technologies and Internet Commerce (CIMCA-IAWTIC'06)}, volume~2,
  pages 308--314. IEEE, 2005.

\bibitem{rios2022security}
Erkuden Rios, Mariv{\'\i} Higuero, Xabier Larrucea, Massimiliano Rak, Valentina
  Casola, and Eider Iturbe.
\newblock Security and privacy service level agreement composition for internet
  of things systems on top of standard controls.
\newblock {\em Computers \& Electrical Engineering}, 98:107690, 2022.

\bibitem{rios2019service}
Erkuden Rios, Eider Iturbe, Xabier Larrucea, Massimiliano Rak, Wissam Mallouli,
  Jacek Dominiak, Victor Munt{\'e}s, Peter Matthews, and Luis Gonzalez.
\newblock Service level agreement-based gdpr compliance and security assurance
  in (multi) cloud-based systems.
\newblock {\em IET Software}, 13(3):213--222, 2019.

\bibitem{rovers1970iso}
Mart Rovers.
\newblock {\em ISO/IEC 20000-1: 2011-A Pocket Guide}.
\newblock Van Haren, 1970.

\bibitem{tawalbeh2020iot}
Lo‚Äôai Tawalbeh, Fadi Muheidat, Mais Tawalbeh, and Muhannad Quwaider.
\newblock Iot privacy and security: Challenges and solutions.
\newblock {\em Applied Sciences}, 10(12):4102, 2020.

\bibitem{theate2021application}
Thibaut Th{\'e}ate and Damien Ernst.
\newblock An application of deep reinforcement learning to algorithmic trading.
\newblock {\em Expert Systems with Applications}, 173:114632, 2021.

\bibitem{theodoridis2013developing}
Evangelos Theodoridis, Georgios Mylonas, and Ioannis Chatzigiannakis.
\newblock Developing an iot smart city framework.
\newblock In {\em IISA 2013}, pages 1--6. IEEE, 2013.

\bibitem{xu2022c}
Yang Xu, Md~Zakirul~Alam Bhuiyan, Tian Wang, Xiaokang Zhou, and Amit~Kumar
  Singh.
\newblock C-fdrl: context-aware privacy-preserving offloading through federated
  deep reinforcement learning in cloud-enabled iot.
\newblock {\em IEEE Transactions on Industrial Informatics}, 19(2):1155--1164,
  2022.

\bibitem{zappatore2015sla}
Marco Zappatore, Antonella Longo, and Mario~A Bochicchio.
\newblock Sla composition in service networks: A tool for representing
  relationships between slas and contracts.
\newblock In {\em Proceedings of the 30th Annual ACM Symposium on Applied
  Computing}, pages 1219--1224, 2015.

\end{thebibliography}

\end{document}